\documentclass[twocolumn,preprintnumbers,nofootinbib,prd,superscriptaddress,aps]{revtex4-1}

\usepackage[utf8]{inputenc} 
\usepackage{graphicx,amssymb,amsmath,amsthm,amsfonts,epstopdf,epsfig,epsf,times}
\usepackage[linktocpage]{hyperref}
\usepackage[usenames]{color}
\usepackage{epstopdf}
\usepackage{textcomp}
\usepackage{bm}
\usepackage{latexsym}
\usepackage{rotating}
\usepackage{hyperref}
\usepackage{color}
\usepackage{longtable}
\usepackage{enumerate}
\usepackage{tensor}
\usepackage{stmaryrd}
\usepackage[normalem]{ulem}
\usepackage{mathtools}
\usepackage{url}
\usepackage{multirow}
\usepackage{graphicx}
\usepackage[caption=false]{subfig}
\usepackage{mathtools}
\usepackage{verbatim}
\usepackage{soul,xcolor}
\usepackage{xspace}

\setstcolor{red}

\makeatletter
\DeclareRobustCommand\onedot{\futurelet\@let@token\@onedot}
\def\@onedot{\ifx\@let@token.\else.\null\fi\xspace}
\def\ie{\emph{i.e}\onedot} 
\makeatother

\definecolor{coolblack}{rgb}{0.0, 0.18, 0.39}
\definecolor{darkred}{rgb}{0.5,0,0}
\definecolor{darkgreen}{rgb}{0,0.5,0}
\definecolor{darkblue}{rgb}{0,0,0.5}
\definecolor{lapislazuli}{rgb}{0.15, 0.38, 0.61}
\definecolor{venetianred}{rgb}{0.78, 0.03, 0.08}
\definecolor{bleudefrance}{rgb}{0.19, 0.55, 0.91}
\definecolor{dogwoodrose}{rgb}{0.84, 0.09, 0.41}
\definecolor{dogwoodrose}{rgb}{0.84, 0.09, 0.41}
\definecolor{darkorgane}{rgb}{1,0.549,0}
\hypersetup{colorlinks=true, citecolor=darkblue, linkcolor=darkblue, 
urlcolor = darkblue}
\definecolor{olive}{rgb}{0.5, 0.5, 0.0}

\newcommand{\ben}{\begin{enumerate}}
\newcommand{\een}{\end{enumerate}}

\def\be{\begin{equation}}
\def\ee{\end{equation}}
\newcommand{\beq}{\begin{eqnarray}}
\newcommand{\eeq}{\end{eqnarray}} 
\newcommand{\ba}{\begin{align}}
\newcommand{\ea}{\end{align}}
\newcommand{\Ricci}{\mathcal{R}}

\def\be{\begin{equation}}
\def\ee{\end{equation}}
	
\newcommand{\bea}{\begin{eqnarray}}
\newcommand{\eea}{\end{eqnarray}}

\def\Lie{\mathcal{L}}

\DeclareMathOperator\diag{diag}

\usepackage{physics}

\begin{document}

\title{Piercing of a boson star by a black hole}

\author{Vitor Cardoso}
\affiliation{Niels Bohr International Academy, Niels Bohr Institute, Blegdamsvej 17, 2100 Copenhagen, Denmark}
\affiliation{CENTRA, Departamento de F\'{\i}sica, Instituto Superior T\'ecnico -- IST, Universidade de Lisboa -- UL,
Avenida Rovisco Pais 1, 1049-001 Lisboa, Portugal}
\author{Taishi Ikeda}
\affiliation{Dipartimento di Fisica, 
``Sapienza'' Universit\'{a} di Roma, Piazzale Aldo Moro 5, 00185, Roma, Italy}
\author{Zhen Zhong}
\affiliation{CENTRA, Departamento de F\'{\i}sica, Instituto Superior T\'ecnico -- IST, Universidade de Lisboa -- UL,
Avenida Rovisco Pais 1, 1049-001 Lisboa, Portugal}
\author{Miguel Zilhão}
\affiliation{Departamento de Matem\'atica da Universidade de Aveiro and
  Centre for Research and Development in Mathematics and Applications (CIDMA),
  Campus de Santiago, 3810-183 Aveiro, Portugal}
\affiliation{CENTRA, Departamento de F\'{\i}sica, Instituto Superior T\'ecnico -- IST, Universidade de Lisboa -- UL,
Avenida Rovisco Pais 1, 1049-001 Lisboa, Portugal}

\begin{abstract} 
New light fundamental fields are natural candidates for all or a fraction of dark matter. Self-gravitating structures of such fields might be common
objects in the universe, and could comprise even galactic halos. These structures would interact gravitationally with black holes, a process of the utmost importance since it dictates their lifetime,
the black hole motion and possible gravitational radiation emission. 
Here, we study the dynamics of a black hole piercing through a much larger fully relativistic boson star, made of a complex minimally coupled massive scalar without self-interactions. As the black hole pierces through the bosonic structure, it is slowed down by accretion and dynamical friction, giving rise to gravitational-wave emission. Since we are interested in studying the interaction with large and heavy scalar structures, we consider mass ratios up to $q\sim 10$ and length ratios ${\cal L} \sim 62$.
Somewhat surprisingly, for all our simulations, the black hole accretes more than 95\% of the boson star material, even if an initially small black hole collides with large velocity.
This is a consequence of an extreme ``tidal capture'' process, which binds the black hole and the boson star together, for these mass ratios.
We find evidence of a ``gravitational atom'' left behind as a product of the process.
\end{abstract}

\maketitle

\section{Introduction}\label{Introduction}
The nature of the matter making up most of the universe is unknown.
There is overwhelming evidence for the existence of dark matter (DM) of an unknown nature and properties, which nevertheless interacts gravitationally~\cite{Freese:2008cz,Navarro:1995iw,Clowe:2006eq,Bertone:2004pz}. Efforts to determine the properties of such matter and to place it in a theoretical framework have so far been unsuccessful, but will continue vigorously for years to come~\cite{Kahlhoefer:2017dnp,PerezdelosHeros:2020qyt}. If the standard model of particle physics is a good guide, one can expect new ``dark stars'' of various types (depending on the number and properties of the putative new elementary particles passing as dark matter), which can make up a significant fraction of astrophysical environments. The scales at which such new structures will appear depend, in particular on the fundamental scale dictated by the fundamental constituents of the, hitherto invisible, new fields.

Here we entertain the possibility that there are new fundamental, scalar degrees of freedom minimally coupled to gravity, and that these form localized, self-gravitating objects. 
It is well known that for complex scalar fields---such as the ones we focus on---boson stars (BSs) form rather generically as a consequence of gravitational collapse~\cite{Kaup:1968zz,Ruffini:1969qy,Liebling:2012fv,Okawa:2013jba,Brito:2015yga}. For real scalars, similar objects exist and form (see Refs.~\cite{Liebling:2012fv,Brito:2015yfh} and work cited therein; the extension to vector degrees of freedom can also be considered~\cite{Brito:2015pxa}). In the absence of self-interactions, the maximum mass $M_{\rm max}$ of such configurations is dictated by the mass of the fundamental boson $\mu$, as $M_{\rm max}=0.8M_{\odot}\times (10^{-10}\,{\rm eV}/\mu)$ \cite{Seidel:1990jh}.
For sufficiently light fields, BSs can therefore have stellar masses, or even galactic-scale masses. Indeed, there are indications that such solutions describe well dark matter cores in halos.  These models are often referred to as fuzzy DM models, and require ultralight bosonic fields (we refer the reader to Refs.~\cite{Robles:2012uy,Hui:2016ltb,Bar:2019bqz,Bar:2018acw,Desjacques:2019zhf,Davoudiasl:2019nlo,Annulli:2020lyc,Schive:2014dra}, but the literature on the subject is very large and growing).

Dark stars have so far gone undetected, but the advent of gravitational-wave (GW) astronomy may also mark the beginning of the illumination of such dark components of our Universe~\cite{Barack:2018yly,Cardoso:2019rvt,Giudice:2016zpa,Ellis:2017jgp}. Understanding the behavior of dark matter when moving perturbers drift by, or when a binary inspirals within a DM medium
is crucial for attempts at detecting DM via GWs. In the presence of a nontrivial environment accretion, gravitational drag and the self-gravity of the medium contribute to a small, but potentially observable, change of the GW phase~\cite{Eda:2013gg,Macedo:2013qea,Barausse:2014tra,Hannuksela:2018izj,Cardoso:2019rou,Baumann:2019ztm,Kavanagh:2020cfn,Annulli:2020lyc,Zwick:2021dlg,Vicente:2022ivh}.

When the length scales between different objects are similar, numerical relativity can be used to extract accurate predictions for the dynamics of the objects and of gravitational waveforms~\cite{Palenzuela:2017kcg,Bustillo:2020syj,Bezares:2022obu,Cardoso:2014uka}. When the scales are too different, one needs to rely on other methods.
The tidal deformability of BSs leaves an imprint in gravitational waveforms and was considered recently~\cite{Bustillo:2020syj,Cardoso:2017cfl,Sennett:2017etc}.
Dynamical friction in scalar structures was also studied recently and allows us to understand the slowdown of bodies moving within scalar structures~\cite{Hui:2016ltb,Annulli:2020lyc,Traykova:2021dua,Vicente:2022ivh}. 
The evolution of a compact binary within a large scale BS was studied within a pointlike approximation for the binary~\cite{Annulli:2020lyc}, and it predicts a $-6$-PN
dephasing effect, potentially observable.

Here, we wish to bridge the gap between these two types of results, and consider the motion of a small black hole (BH) as it ``pierces'' a large BS structure
fully nonlinearly.
The small BH will be subjected to friction, it accretes a portion of the scalar material from the BS, and it emits GWs carrying energy and momentum. Likewise, as a consequence of the BH motion, the BS itself will move and be nonlinearly perturbed---or even destroyed entirely.

We use units where the speed of light, Newton's constant and reduced Planck's constant are all set to unity, $c=G=\hbar=1$.

\section{Boson star construction}
\label{sec:BS-eom}
We consider a minimally coupled, complex scalar field $\Phi$ described by the Einstein-Klein-Gordon action,
\[
S=\int d^{4} x \sqrt{-g}\left[\frac{\Ricci}{16\pi}-\left(g^{ab} \nabla_a\Phi \nabla_b\Phi^{*}+\mu^{2} \Phi \Phi^{*}\right)\right]\,,
\]
where $g_{ab}$ is the spacetime metric, $\Ricci$ is the Ricci scalar, $\mu$ is the mass of the scalar field and
$^*$ denotes complex conjugation. From the action, the equations of motion are
\begin{align}
&\Ricci_{ab} - \frac{1}{2} \Ricci g_{ab} = 8\pi T_{ab}\,,\\
&g^{ab}\nabla_a\nabla_b\Phi = \mu^2\Phi\,,
\end{align}
with the stress-energy tensor
\begin{equation}
T^{ab}=\nabla^a\Phi \nabla^b\Phi^{*}+\nabla^a\Phi^{*}\nabla^b\Phi-g^{ab}\left(\nabla^c\Phi \nabla_c\Phi^{*}+\mu^{2} \Phi \Phi^{*}\right)\,.
\end{equation}

Following Ref.~\cite{Liebling:2012fv}, we derive equilibrium equations in spherical symmetry by assuming a harmonic ansatz for the scalar field and a stationary geometry,
\begin{align}
\Phi(\mathbf{r}, t)&=\phi(\mathbf{r}) e^{i \omega t}\,,\label{harmonic_ansatz}\\
ds^2&=-\alpha(r)^{2} dt^2+a(r)^{2} dr^2+r^{2} d\Omega_2^2\,.\label{metric}
\end{align}
Then, the spherically symmetric Einstein-Klein-Gordon system can be written as three ordinary differential equations
\begin{align*}
&a'=\frac{a}{2}\left\{\frac{1-a^{2}}{r}+8 \pi  r\left[\left(\frac{\omega^{2}}{\alpha^{2}}+\mu^{2}\right) a^{2} \phi^{2}+(\phi')^{2}\right]\right\}\,,\\
&\alpha'=\frac{\alpha}{2}\left\{\frac{a^{2}-1}{r}+8 \pi  r\left[\left(\frac{\omega^{2}}{\alpha^{2}}-\mu^{2}\right) a^{2} \phi^{2}+(\phi')^{2}\right]\right\}\,,\\
&\phi''=-\left\{1+a^{2}-8 \pi r^{2} a^{2} \mu^{2} \phi^{2}\right\} \frac{\phi'}{r}-\left(\frac{\omega^{2}}{\alpha^{2}}-\mu^{2}\right) \phi a^{2}\,,
\end{align*}
where primes stand for radial derivatives. In order to obtain a physical solution, the following boundary conditions must be imposed on this system.
\begin{subequations}
\begin{align}
& \phi(0) = \phi_0\,,\qquad \phi'(0) = 0\,,\qquad a(0)= 1\,,\label{bcPhiA}\\
& \lim_{r\to\infty}\phi(r)= 0\,,\label{bcPhi}\\
& \lim_{r\to\infty} \alpha(r) a(r) = 1\,.\label{bcAlphaA}
\end{align}
\end{subequations}
Here, $\phi_0$ can be specified arbitrarily, and it roughly determines the mass of the boson star. We can find a simpler system by rescaling the variables in the following manner,
\[
\tilde{\phi} \equiv \sqrt{8 \pi} \phi, \quad \tilde{r} \equiv \mu r, \quad \tilde{t} \equiv \omega t, \quad \tilde{\alpha} \equiv(\mu / \omega) \alpha\,.
\]
Then the equations become
\begin{equation}
  \label{eq:BS-system}
\begin{aligned}
  &a'=\frac{a}{2}\left\{\frac{1-a^{2}}{\tilde{r}}+ \tilde{r}\left[\left(\frac{1}{\tilde{\alpha}^{2}}+1\right) a^{2} \tilde{\phi}^{2}+(\tilde{\phi}')^{2}\right]\right\}\,,\\
  &\tilde{\alpha}'=\frac{\tilde{\alpha}}{2}\left\{\frac{a^{2}-1}{\tilde{r}}+ \tilde{r}\left[\left(\frac{1}{\tilde{\alpha}^{2}}-1\right) a^{2} \tilde{\phi}^{2}+(\tilde{\phi}')^{2}\right]\right\}\,,\\
  &\tilde{\phi}''=-\left\{1+a^{2}-\tilde{r}^{2} a^{2} \tilde{\phi}^{2}\right\} \frac{\tilde{\phi}'}{\tilde{r}}-\left(\frac{1}{\tilde{\alpha}^{2}}-1\right) \tilde{\phi} a^{2}\,.
\end{aligned}
\end{equation}
where primes stand for the derivatives with respect to $\tilde{r}$. Note that both $\mu$ and $\omega$ drop out of the equations. We will use the mass parameter $\mu$ as our unit.
%

To integrate these equations, we need to understand their asymptotic behavior. At the origin, $\tilde{r} = 0$, we can expand all quantities in a Taylor series 
to find
\begin{align*}
  a(\tilde{r})&=1 + \frac{\tilde{r}^2(\tilde{\alpha}_0^2 + 1)\tilde{\phi}_0^2}{6\tilde{\alpha}_0^2} + \mathcal{O}(\tilde{r}^4)\,,\\
  \tilde{\alpha}(\tilde{r})&=\tilde{\alpha}_0 - \frac{\tilde{r}^2(\tilde{\alpha}_0^2 - 2)\tilde{\phi}_0^2}{6\tilde{\alpha}_0} + \mathcal{O}(\tilde{r}^4)\,,\\
  \tilde{\phi}(\tilde{r})&= \tilde{\phi}_0 + \frac{\tilde{r}^2 (\tilde{\alpha}_0^2 - 1) \tilde{\phi}_0}{6 \tilde{\alpha}_0^2} + \mathcal{O}(\tilde{r}^4)\,,
\end{align*}
where $\tilde{\phi}(0) = \tilde{\phi}_0$, $\tilde{\alpha}(0) = \tilde{\alpha}_0$.
Notice that the form of the metric in Eq.~\eqref{metric} is consistent with the Schwarzschild metric at large distances where
\begin{equation}
a(r) = \left(1-\frac{2M}{r}\right)^{- \frac{1}{2}}\,,\label{aMeqn}
\end{equation}
with $M$ the ADM mass of the spacetime. This allows us to define a more general mass aspect function
\begin{equation}
M(r)=\frac{r}{2}\left(1-\frac{1}{a^{2}(r)}\right)\,,
\end{equation}
which measures the total mass contained in a sphere of radius $r$. Furthermore, we can use boundary condition Eq.~\eqref{bcAlphaA} together with Eq.~\eqref{aMeqn} to get $\alpha(r)$ at large distances
\begin{equation}
\alpha(r) = \left(1 - \frac{2M}{r}\right)^{\frac{1}{2}}\,.
\end{equation}
We are now ready to solve for the boson star structure using a shooting method to solve Eqs.~\eqref{eq:BS-system}: we fix $\tilde{\phi}_0$ and shoot for $\tilde{\alpha}_0$ using boundary conditions Eqs.~(\ref{bcPhiA}) and (\ref{bcPhi}), so as to have an asymptotically flat spacetime. There are many possibilities for $\tilde{\alpha}_0$, which correspond to different excited states of the boson star.
Once we have solved for $a$, $\tilde \alpha$ and $\tilde \phi$ we can recover $\omega$ by using Eq.~(\ref{bcAlphaA}). In practice, since we use a finite grid, we use the values of $\alpha$ and $a$ at the last radial grid point of the solution we have (see below for details on how good the agreement is with the Schwarzschild metric at $r \sim 100$).

Finally, to perform numerical evolutions in the absence of symmetries, isotropic coordinates are preferred.
In isotropic coordinates, the spherically symmetric metric can be written as~\cite{Liebling:2012fv}
\begin{equation}
ds^{2}=-\alpha(R)^{2} dt^{2}+\psi(R)^{4}\left(d R^{2}+R^{2} d \Omega^{2}\right)\,,
\end{equation}
where $\psi$ is the conformal factor. A change of the radial coordinate $R = R(r)$ can transform the solution obtained in Schwarzschild coordinates into isotropic ones, in particular,
%
\begin{equation}
\frac{d R}{d r} =a \frac{R}{r}\,.\label{dR_eqn}
\end{equation}
%
%
In a boson star the scalar field decays exponentially, and therefore the solution quickly asymptotes to a Schwarzschild exterior. We can thus integrate Eq.~\eqref{dR_eqn} by imposing a Schwarzschild exterior of mass $M$ at large distances, with
\begin{equation}
r(R) = M + \frac{M^2}{4R} + R \,.\label{rR_eqn}
\end{equation}

We perform the coordinate transformation \eqref{dR_eqn} numerically where Eq.~\eqref{rR_eqn} is used when $r > r_{\mathrm{max}}$. We solve this equation via a shooting method, where we integrate outwards; we therefore
need to understand the behavior $R(r)$ for small $r$. Taylor expanding at $\tilde{r} = 0$, \ie., $R(r) = \sum_n R^{(n)}r^n/n! $, we find
\begin{equation}
R(r) = c r - \frac{c r^3(1+\alpha_0^2)\phi_0^2}{12 \alpha_0^2} + \mathcal{O}(r^5)\,.
\end{equation}
Then, $R'(0) = c$ is a free parameter which we determine by shooting to the exterior solution \eqref{rR_eqn}.
In practice we set $r_{\mathrm{max}} \sim 100$ and we can verify that we effectively recover the Schwarzschild metric from this point onward:
$|\phi| < 10^{-16}$, $|a(r) - \left(1-2M/r\right)^{- \frac{1}{2}}| < 10^{-16}$.
Hence, our solution is smoothly connected.

\begin{figure}[th]
  \includegraphics[width=0.45\textwidth]{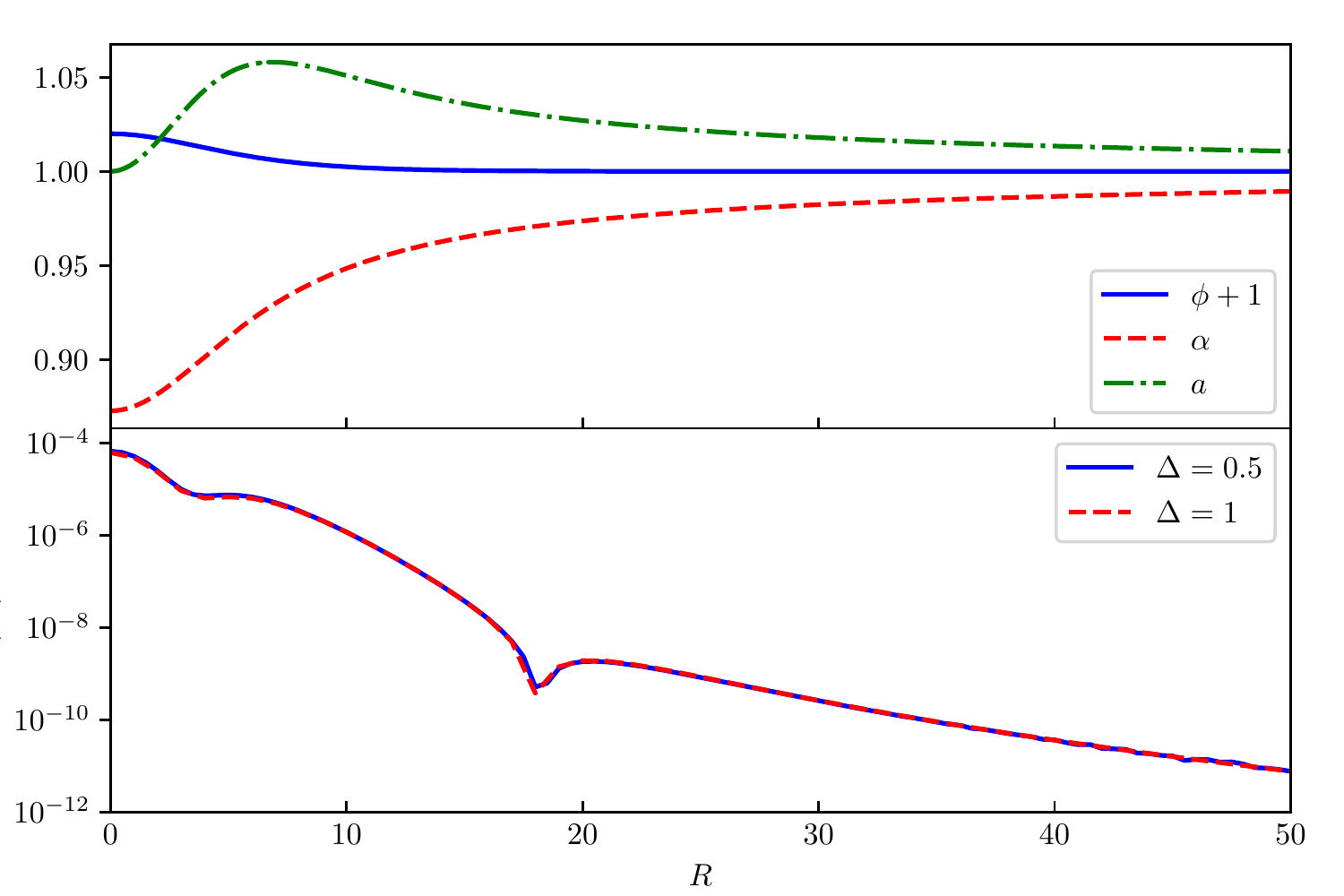}
  \caption{Configuration describing an isolated BS with mass $M=0.53$ in isolation, corresponding to a value of the scalar field at the center $\phi_0=0.02,\,\alpha_0=0.873,\,\omega=0.936$.
  {\bf Top:} scalar field and metric components as a function of the radial coordinate.
  {\bf Bottom:} Hamiltonian constraint along the $R$ axis. The red curve has been multiplied by 16, the expected factor for fourth-order convergence. We use mesh refinement, and $\Delta$ represents $\dd{x}, \dd{y}$, and $\dd{z}$ of the coarsest level.
   \label{fig:bs_isolation}}
\end{figure}
In the following, we use a ground-state BS with mass $M=0.53$ in isolation, corresponding to BS parameters $\mu = 1,\,\phi_0=0.02,\,\alpha_0=0.873,\,\omega=0.936$. The solution is summarized in the top panel of Fig.~\ref{fig:bs_isolation}. Note that $R_{98} = 12.39$ is the radius of the isolated BS solution which encloses 98\% of the BS mass. A measure of the correctness of this solution can be assessed by the violation of the Hamiltonian constraint (introduced in the next section, see Eq.~\eqref{eq:Hamiltonian}), shown in the bottom panel of Fig.~\ref{fig:bs_isolation}. Notice that when the resolution increases, the constraint violation decreases in accordance with the expected fourth-order convergence of our results (we use fourth-order accurate finite-difference operators throughout).

When the BS is dilute, the relevant equations governing the structure of isolated BSs reduce to the Schrodinger-Poisson
equations~\cite{Chavanis_2011,Annulli:2020lyc}. In this limit, all BS solutions are related via simple scaling relations, and accurate fits are given by 
expressions (45)-(50) in Ref.~\cite{Annulli:2020lyc}. In this regime, the radius $R_{98}$ (defined by the areal radius containing 98\% of the BS mass) can be related to the total mass via
$M\mu\simeq 9.1/R_{98}\mu$. The BS that we use as a reference is not fully within the Newtonian regime.

\section{Dynamical BH-BS spacetimes}
\subsection{3+1 decomposition and evolution procedure}

Our strategy to perform numerical evolutions makes use of the standard 3+1
decomposition~\cite{Gourgoulhon:2007ue,Alcubierre08a}, whereby a 3-metric
$\gamma_{ab}$ is introduced via
\begin{equation}
\gamma_{ab} = g_{ab} + n_{a} n_{b} \,,  \label{eq:3metric}
\end{equation}
where $n^{a}$ denotes a unit timelike vector normal to a spacelike hypersurface. 
The full spacetime metric $g_{ab}$ can then be written in the form
\begin{align}
ds^{2} & = g_{ab} dx^{a} dx^{b} \notag \\
       & = - \left( \alpha^{2} - \beta^{i} \beta_{i} \right) \, dt^{2}
             + 2 \beta_{i} \,dt \,dx^{i}
             +   \gamma_{ij} \, dx^{i} \,dx^{j}\,, \label{eq:LineElement}
\end{align}
where $\alpha$ and $\beta$ are the usual lapse and shift gauge functions, and the indices $i,j,\dots$ run from 1 to 3.

To write the evolution equations in this formalism we introduce the extrinsic curvature $K_{ij}$,
\begin{equation}
K_{ij}  =   - \frac{1}{2\alpha} \left( \partial_{t} - \Lie_{\beta} \right) \gamma_{ij} \,,\label{eq:KijDef}
\end{equation}
and the ``conjugate momentum'' of the complex scalar field $\Phi$,
\begin{equation}
K_{\Phi} = -\frac{1}{2\alpha}  \left( \partial_{t} - \Lie_{\beta} \right) \Phi \,,\label{eq:Kphi}
\end{equation}
where $\Lie$ denotes the Lie derivative.
The evolution equations then take the form
%
%
\begin{subequations}
\begin{align}
  \partial_{t} \gamma_{ij} & = - 2 \alpha K_{ij} + \Lie_{\beta} \gamma_{ij} \,,
                       \label{eq:dtgamma} \\
  \partial_{t} K_{ij}      & =  - D_{i} \partial_{j} \alpha
                       + \alpha \left( ~^{3}\Ricci_{ij} - 2 K_{ik} K^{k}{}_{j} + K K_{ij} \right) \nonumber \\
                       & \quad + \Lie_{\beta} K_{ij} 
                       + 4\pi \alpha \left[ (S-\rho) \gamma_{ij} - 2 S_{ij} \right] \,,
                                         \label{eq:dtKij} \\
  \partial_{t} \Phi & = - 2 \alpha K_\Phi + \Lie_{\beta} \Phi\
                \,, \label{eq:dtPhi} \\
  \partial_{t} K_\Phi &  = \alpha \left( K K_{\Phi} - \frac{1}{2} \gamma^{ij} D_i \partial_j \Phi
                  + \frac{1}{2} \mu^2 \Phi \right) \nonumber \\
                 & \quad - \frac{1}{2} \gamma^{ij} \partial_i \alpha \partial_j \Phi
                       + \Lie_{\beta} K_\Phi \,, \label{eq:dtKphi}
\end{align}
\end{subequations}
%
which is subject to the following constraints
\begin{align}
\label{eq:Hamiltonian}
\mathcal{H} & \equiv~^{3}\Ricci - K_{ij} K^{ij} + K^2 - 16 \pi \rho
       = 0\,,\\
\label{eq:momentumConstraint}
\mathcal{M}_{i} & \equiv D^{j} K_{ij} - D_{i} K 
        - 8\pi j_i
       = 0 \,.
\end{align}
where  $K \equiv \gamma^{ij} K_{ij}$,
$D_i$ and $^{3}\Ricci_{ij}$ denote, respectively, the covariant derivative and Ricci tensor with respect to the $3$-metric, and the source terms are given by
\begin{equation}
  \label{eq:source}
   \begin{aligned}
  \rho & \equiv T^{a b}n_{a}n_{b} \,,\\
  j_i  &\equiv -\gamma_{ia} T^{a b}n_{b} \,, \\
  S_{ij} &\equiv \gamma^{a}{}_i \gamma^{b}{}_j T_{a b} \,, \\
  S     & \equiv \gamma^{ij}S_{ij} \,.
   \end{aligned}
\end{equation}

For the numerical simulations we rewrite the evolution equations above in the
Baumgarte-Shapiro-Shibata-Nakamura form~\cite{Nakamura87,Shibata:1995we,Baumgarte:1998te}, as detailed
in Ref.~\cite{Cunha:2017wao}, and use the Einstein~Toolkit
infrastructure~\cite{Loffler:2011ay,Zilhao:2013hia,EinsteinToolkit:2019_10} for
the evolutions. We use Carpet~\cite{Schnetter:2003rb} for mesh refinement
capabilities, AHFinderDirect~\cite{Thornburg:1995cp,Thornburg:2003sf} for
finding and tracking apparent horizons, and
QuasiLocalMeasures~\cite{Dreyer:2002mx} for extracting BH mass. The spacetime metric and scalar field variables are evolved in time using the LeanBSSNMoL and ScalarEvolve codes, which are freely available as
part of the Canuda library~\cite{Canuda_zenodo_3565474}.

For our simulations we use two refinement centers, one center corresponding to the location of the BH and the other to the location of the BS, which sits initially at the origin of the numerical domain.
We use up to $10$ refinement levels for the BH and $3$ refinement levels for the BS.
We always use at least 25 points to cover the BH, thus ensuring enough grid points for an acceptable resolution.
During the evolution, the mesh refinement around the BH moves, and the mesh refinement resolving BSs are fixed at the origin of the numerical domain.
To avoid redundant calculations, we assume reflection symmetry on the $x=0$ and $y=0$ planes (the collision will always be along the $z$ axis).

Throughout the code, derivatives are approximated using fourth-order-accurate
finite-differencing stencils but there are also lower-order elements in the code,
such as prolongation operations, which are only second-order accurate in time.
We use the method of lines with Runge-Kutta 4 to evolve the equations in time
with outgoing (radiative) boundary conditions
and the usual $1+\log$ and Gamma-driver gauge conditions~\cite{Alcubierre08a}.

\subsection{Diagnostics\label{subsec:diagnostics}}
To understand and characterize some of the physics more precisely, we monitor the scalar field around the moving BH, in a frame comoving with the BH.

We simply consider a sphere with constant coordinate radius $\bar{r}$ around the BH, as measured with the numerical coordinates introduced above, 
and we extract the multipole mode of the scalar field on the sphere, which is defined as
\be
\phi_{lm}(t,\bar{r})=\int_{S_{\rm BH}} d^{2}\Omega Y_{lm}^{*}(\Omega)\Phi(t,\bar{r},\Omega)\,,
\ee
where $S_{\rm BH}$ is the sphere around the BH with radius $\bar{r}$. For the small Lorentz boosts considered in this work, such a sphere is also a constant-radius sphere in the BH frame. 

In addition, we monitor the energy $E^{\rm rad}$ and momentum $P^{\rm rad}$ radiated in GWs, which can be be calculated as~\cite{Berti:2007fi},
\begin{align}
  \frac{d E^{\rm rad}(t)}{dt} &=\lim _{r \rightarrow \infty}\left[ \frac{r^{2}}{16 \pi} \int_{\Omega}\left|\int_{-\infty}^{t} \Psi_{4} d \tilde{t}\right|^{2} d \Omega\right], \\
  \frac{d P_i^{\rm rad}(t)}{dt} &=-\lim _{r \rightarrow \infty}\left[\frac{r^{2}}{16 \pi} \int_{\Omega} \ell_{i}\left|\int_{-\infty}^{t} \Psi_{4} d \tilde{t}\right|^{2} d \Omega\right],
\end{align}
where $\ell_{i}=(-\sin \theta \cos \phi, -\sin \theta \sin \phi, -\cos \theta)$, and $\Psi_{4}$ is the Newman-Penrose scalar, which is defined in Appendix C of Ref.~\cite{Sperhake:2006cy}.

Besides, we also monitor the energy density and momentum flux of the scalar field into the BH horizon. For any vector field, we have
\begin{equation}
Q =\int_{\Omega} d^{3}x\alpha\sqrt{\gamma}T_{a}^{~t}\xi^{a}\,,\label{eq_defQ}
\end{equation}
where $T_{ab}$ is the energy-momentum tensor, $\gamma$ is the determination of the 3-metric, and $\alpha$ is the lapse function.
When $\xi^a = (\pdv{t})^a$ we denote $Q$ by $Q_t$, and for $\xi^a = (\pdv{z})^a$ we denote $Q$ by $Q_z$. Even though $\xi$ is not a Killing vector, the charge defined in Eq.~\eqref{eq_defQ} is a good measure of the scalar field energy, and it has been adopted by other authors (e.g. Ref.~\cite{East:2017mrj}).

\subsection{Initial data}

For our evolution procedure, the relevant initial data amount to specifying the spatial profile of the 3-metric, extrinsic curvature, lapse and shift, as well as the scalar field, at a given time slice.
We have described in Sec.~\ref{sec:BS-eom} our construction of stationary BS spacetimes. Isolated BH spacetimes are known analytically, and to construct
spacetimes containing both objects we superpose these two solutions using a procedure analogous to the one outlined in Ref.~\cite{PhysRevD.78.101501}, which we can summarize as follows. The spacetime is described by
\begin{subequations}
\begin{align}
  K_{ij} &= K_{ij}^{\mathrm{(BH)}} + K_{ij}^{\mathrm{(BS)}} + \delta K_{ij}\,,\\
  \psi &= \psi^{\mathrm{(BH)}} + \psi^{\mathrm{(BS)}} - 1 + \delta\psi\,,\\
  \gamma_{ij} &= \psi^4\diag(1,1,B^2) \,, \label{eqn:metric_gamma} \\
  B^2 &= \Gamma^2 \left[ 1 - v^2 (3 - \psi^{\mathrm{(BH)}} + \psi^{\mathrm{(BS)}})^2 \psi^{-6} \right] \,,
\end{align}
\end{subequations}
where $\Gamma \equiv 1 / \sqrt{1 - v^2}$, $v$ is the speed of the BH, $K_{ij}^{\mathrm{(BH)}}$ and $K_{ij}^{\mathrm{(BS)}}$ are the extrinsic curvatures of the BH and BS respectively, while $\psi^{\mathrm{(BH)}}$ and $\psi^{\mathrm{(BS)}}$ are the conformal factors of the BH and BS solutions, respectively. In the above system $\delta K_{ij}$ and $\delta \psi$ are correction terms, which should be solved for by solving the appropriate elliptic system. For simplicity, here we set $\delta K_{ij} = 0$ and $\delta \psi = 0$. This means that these initial data do not satisfy the constraint equations (\ref{eq:Hamiltonian}) and (\ref{eq:momentumConstraint}). While not ideal, the constraint violation incurred is small for large initial distances, and such a construction has been standard practice in studies of BS binaries~\cite{Sanchis-Gual:2018oui,Sanchis-Gual:2020mzb,Jaramillo:2022zwg}.

One can find explicit expressions for the metric and extrinsic curvature of a boosted BH in Ref.~\cite{PhysRevD.78.101501}.
For the spherically symmetric BS solutions that we consider, $K^{\mathrm{(BS)}}_{ij} = 0$ and
\[
  K^{\mathrm{(BS)}}_{\Phi}
	= -\frac{1}{2\alpha} \partial_{t} \Phi = -\frac{i\omega\phi}{2\alpha}\,,
\]
where $\alpha$ is the lapse function obtained previously for BSs in isolation; $\omega$ and $\phi$ are defined in Eq.~(\ref{harmonic_ansatz}). 
Finally, we fix the initial conditions for the gauge variables by choosing $\alpha = \psi^{-2}$, and $\beta^i = 0$.

We place the BS at the center of our coordinates, and we keep its parameters fixed in all our simulations: as we discussed above, we consider a BS which in isolation is characterized by a total mass $M=0.53$, corresponding to a value of the scalar field at the center, $\phi_0=0.02,\,\alpha_0=0.873,\,\omega=0.936$, as shown in Fig.~\ref{fig:bs_isolation}. Hereafter we fix units where $\mu=1$---all our results will be shown and discussed in these units.

\begin{table}[htb]
\caption{List of simulations analyzed for collisions between a BH of mass parameter $M_{\rm BH}$ and a BS of mass $M=0.53$.
The BH has initial velocity $v_0$ along the $z$ axis, and starts from position $z_0$. The BS is characterized by a frequency $\omega=0.936$ and values at the origin $\phi_0=0.02,\,\alpha_0=0.873$. The total energy is $M_{\rm tot}=\Gamma M_{\rm BH}+M-\Gamma M_{\rm BH}M/z_0$ with $\Gamma$ the Lorentz factor. The total momentum of the boosted BH is $\Gamma M_{\rm BH} v_0$. We define a mass ratio $q=M / M_{\rm BH}$ and a length ratio ${\cal L}=R_{98} /(2M_{\rm BH})$. Notice that at the initial time the mass parameter $M_{\mathrm{BH}}$ is equal to the irreducible mass $M_{\mathrm{irr}}$ within better than 0.5\% and the irreducible mass is given by $\mathcal{A} = 16 \pi M_{\mathrm{irr}}^2$, where $\mathcal{A}$ is the area of the apparent horizon.
\label{table:simulations1}}
\begin{ruledtabular}
\begin{tabular}{l@{\hskip 0.1in}c@{\hskip 0.1in}c@{\hskip 0.1in}c@{\hskip 0.1in}c@{\hskip 0.1in}c@{\hskip 0.1in}c@{\hskip 0.1in}c@{\hskip 0.1in}}
Run          & $M_{\rm BH}$&${\cal L}$  & $v_0$      &$z_0$    & $M_{\rm tot}$ &$P_{\rm tot}$ 
\\
\hline
\texttt{IA}  &  1          &6          & $10^{-4}$   &-50      &1.52           & 0            
\\
\texttt{IB}  &  1          &6         & $0.5$       &-200      &1.68           & 0.577        
\\
\texttt{IIA} & 0.4         &16        & $10^{-4}$   &-50       &0.93           & 0             
\\
\texttt{IIB} & 0.4         &16        & $0.5$       &-200      &0.99           & 0.231        
\\
\texttt{IIIA} & 0.2        &31        & $10^{-4}$  &-50        &0.73           & 0            
\\
\texttt{IIIB} & 0.2        &31        & $0.5$      &-200       &0.76           & 0.115        
\\
\texttt{IVA} & 0.1         &62        & $10^{-4}$   &-50       &0.63           & 0            
\\
\texttt{IVB} & 0.1         &62        & $0.5$      &-200       &0.65           & 0.086        
\end{tabular}
\end{ruledtabular}
\end{table}
What we vary in the initial data are the BH parameters, in particular, its initial velocity and mass.
We have studied a variety of initial conditions, summarized in Table~\ref{table:simulations1}. 
Our initial data include a BS-BH mass ratio $q=M/M_{\rm BH}$ ranging from $0.2$ to $2$, and a length ratio ${\cal L}=R_{98}/(2M_{\rm BH})$ from $6$ to $62$.
These ratios are far from those expected for galactic halos~\cite{Annulli:2020lyc}, but they are the only ones possible with current infrastructure. They could be appropriate to describe the interactions between BHs and DM roaming ``lumps,'' or even as a starting point to understand how to extrapolate to other, more extreme ratios.

We also consider two different initial BH velocities. One describes slow infalls, for which we take an almost static BH with $v_{0}=10^{-4}$ at $z_{0}=-50$.
We also consider high velocity collisions, for which $v_{0}=0.5$ at $z_{0}=-200$.
In all cases, the BH is initially well outside the BS, $z_0\gg R_{98}$.

\section{Results}
%
\begin{figure*}[tphb]
  \includegraphics[width=\textwidth]{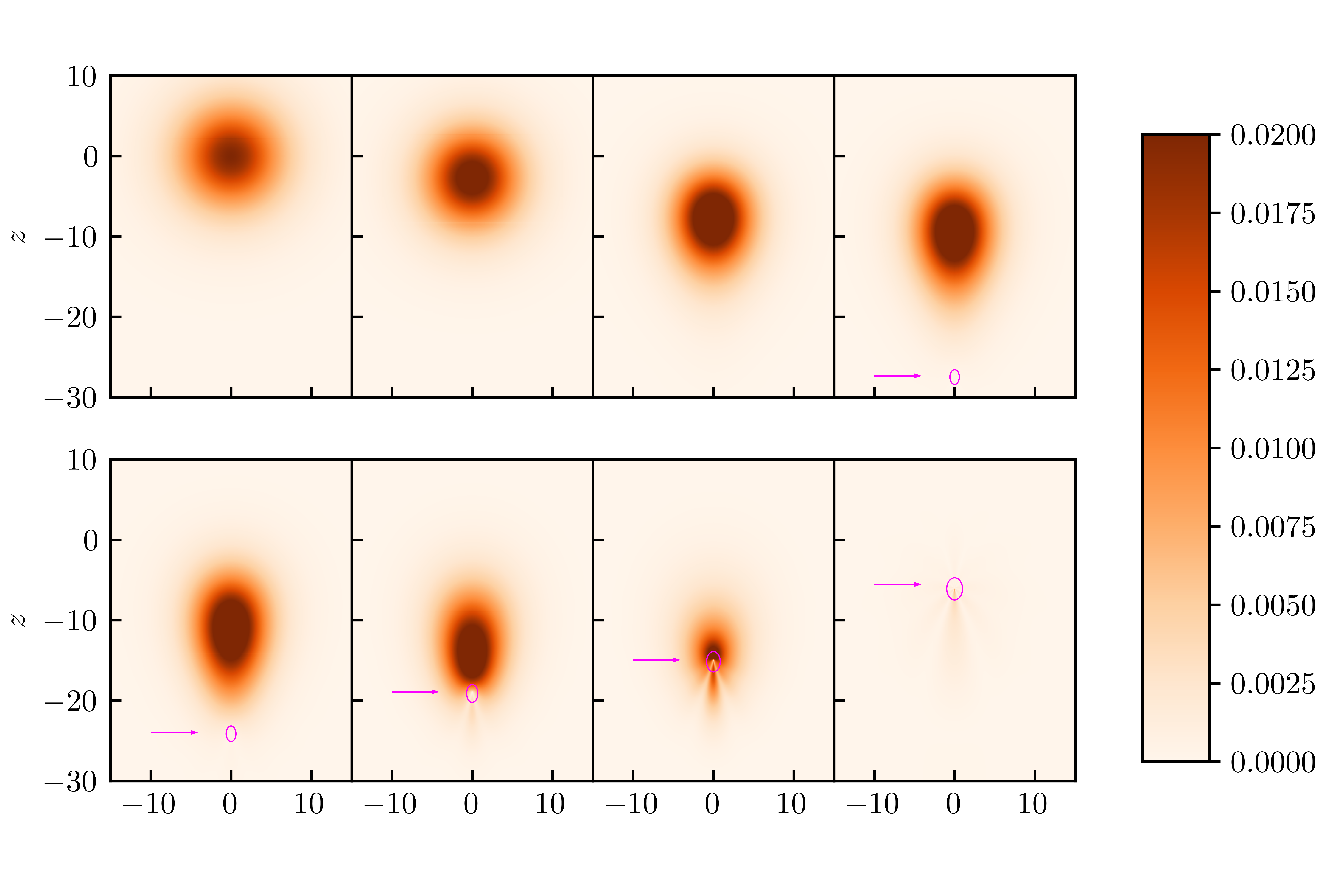}
  \caption{Snapshots of evolution, depicting the scalar field absolute value $|\Phi|$ for the initial data \texttt{IB} in Table~\ref{table:simulations1}. From left to right, the snapshots are taken at instants $t=0.0,\,300.8,\,380.8,\,390.4$ in the top row, and at $t=396.8,\,406.4,\,416.0,\,448.0$ in the bottom row. Pink lines depict contours of constant lapse function $\alpha = 0.2$, which are a good approximation for the location of the horizon, further indicated by arrows. Notice how the BS is tidally distorted as it approaches the BH and how it eventually is almost totally swallowed up by the BH.
Animations are available online~\cite{GRIT_animations}, and also as ancillary files in this submission.
    \label{fig:snapshotsIB}}
\end{figure*}
\begin{figure*}[tphb]
  \includegraphics[width=\textwidth]{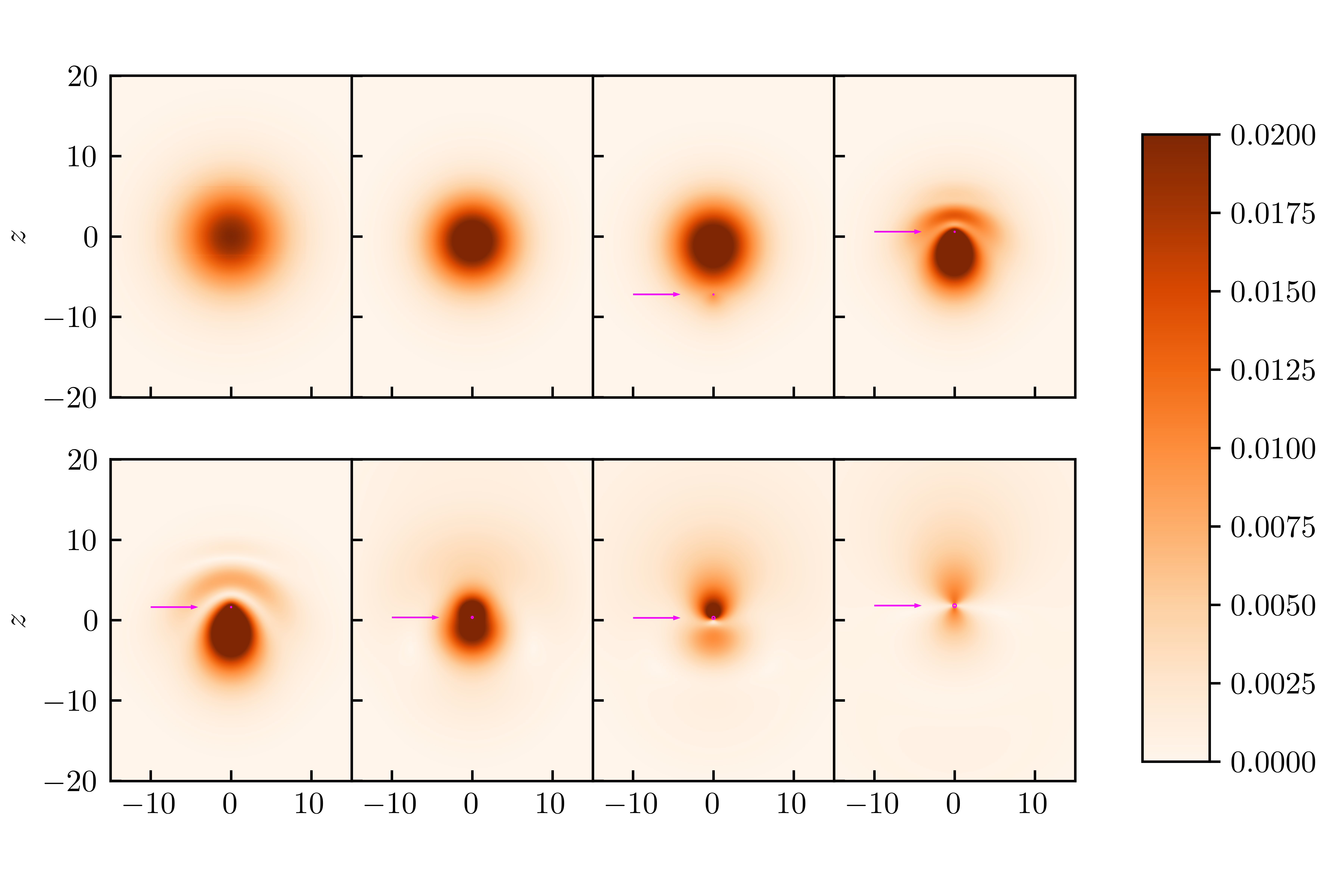}
  \caption{Snapshots of evolution, depicting the scalar field absolute value $|\Phi|$ for the initial data \texttt{IVB} in Table~\ref{table:simulations1}. From left to right, the snapshots are taken at instants $t=0.0,\,349.6,\,389.6,\,408.8$ in the top row, and at $t=415.2,\,436.8,\,450.4,\,480.8$ in the bottom row. Pink lines depict contours of constant lapse function $\alpha = 0.2$, which, as in the previous figure, indicate the horizon location; this region is much smaller (and difficult to see) than the corresponding one of Fig.~\ref{fig:snapshotsIB} since the BH is much smaller. Notice how the BS pulls back the BH during the collision process (panels 5 and 6), and eventually is swallowed up by the BH. Animations are available online~\cite{GRIT_animations}, and also as ancillary files in this submission.
    \label{fig:snapshotsIVB}}
\end{figure*}
%
\subsection{Dynamics and accretion during collision process}

We mostly look at results in the ``lab frame,'' where the BS is at rest, and the BH---initially at $(0, 0, z_0)$---is moving along $z$ axis to eventually collide and interact with it.

\begin{figure}[htb]
  \centering
  \includegraphics[width=0.52\textwidth]{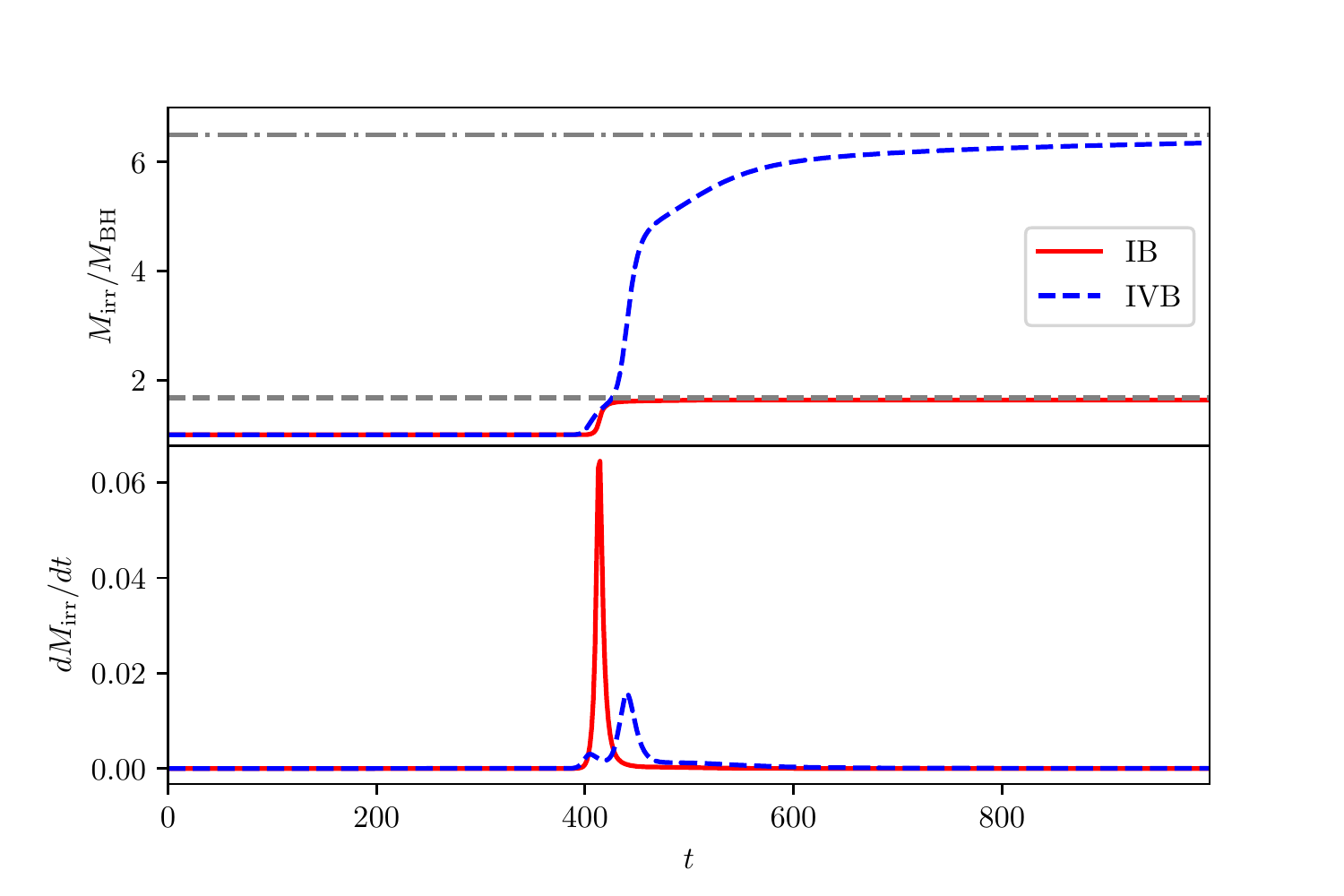}
  \caption{Accretion of scalar onto the BH. {\bf Top panel:} normalized BH irreducible mass $M_{\mathrm{irr}} / M_{\rm BH}$ for simulations \texttt{IB} and \texttt{IVB}. The gray lines are the total mass $M_{\mathrm{tot}}$ normalized by the initial mass $M_{\rm BH}$ given in Table~\ref{table:simulations1}.  At late times the BH mass approaches $M_{\rm tot}$, thus the BH ends up accreting the entire BS. {\bf Bottom panel:} accretion rate for the two different initial data. Notice that there are two accretion stages for simulation \texttt{IVB}, which we argue to be due to tidal effects. The accretion rate is larger than expected for a stationary regime, see main text.
  \label{fig:bh_mass}}
\end{figure}

The numerical convergence of our results is consistent with the discretization scheme we used, as discussed in Appendix~\ref{app:convergence}. The outcome of our numerical simulations are summarized in Figs.~\ref{fig:snapshotsIB}-\ref{fig:gw_flux_500} and Table~\ref{table:simulations_results}.

Snapshots of the scalar field absolute value $|\Phi|=\sqrt{\Phi\Phi^*}$ are shown in Figs.~\ref{fig:snapshotsIB}--\ref{fig:snapshotsIVB}, for initial data \texttt{IB} and \texttt{IVB}. The contour in pink marks the contour of constant lapse function $\alpha=0.2$, which is a good indication of the BH apparent horizon location. It is clear from Fig.~\ref{fig:snapshotsIB} that, in this frame, during the collision the BS moves towards the BH as it approaches. If we define the collision as the instant when the BH crosses $R_{98}$, it is clear from these snapshots that even prior to collision the BS is tidally distorted. Tidal Love numbers of BSs are positive and relatively large as compared to compact systems~\cite{Mendes:2016vdr,Cardoso:2017cfl}, hence the deformation is along the BH-BS axis and visible. The tidal deformation is clear for nearly equal mass objects, as the effects are much stronger in this regime. For the purpose of our study, the case \texttt{IVB} in Fig.~\ref{fig:snapshotsIVB} is more interesting, for it describes a very small BH plunging through a large BS. Tidal effects are now less obvious on length scales of the boson star; however, they become visible and will play a crucial role once the BH pierces through the BS, as we explain below.

It is also evident from the snapshots that the BS is almost entirely swallowed by the BH, which continues moving---albeit with a smaller velocity by momentum conservation---after the interaction with the BS. We find that the BS is nearly totally accreted for {\it all} the initial data in Table~\ref{table:simulations1}; an extreme example concerns simulation \texttt{IVB}: even a BH which is $62$ times smaller than the BS and moving at half the speed of light, ends up ``eating'' all of the BS material. This is seen, in particular, in Fig.~\ref{fig:bh_mass} where we show the BH irreducible mass (estimated with the help of the horizon area) as function of coordinate time. After a relatively short timescale $\sim 1000$, and after the BH has had time to interact with the BS, the final BH mass is at least 95 \% of the total initial energy. We have further quantified accretion by computing the scalar energy in the initial and final spacetime slices [cf. definition~\eqref{eq_defQ}]; our results are shown in Table~\ref{table:simulations_results} and are consistent with total or nearly total accretion of the BS onto the BH, even for such length ratios as ${\cal L}=62$ of run \texttt{IVB}.

Accretion onto a nonmoving BH placed at the center of a Newtonian BS proceeds at a stationary rate $\dot{M}\sim 8\pi \times 10^{-2}(\mu M_{\rm BH})(\mu M_{\rm BS})^4$~\cite{Cardoso:2022zzz}.
For simulation \texttt{IV} this amounts to $\dot{M}\sim 2\times 10^{-4}$, roughly 2 orders of magnitude below what we observe numerically. We can estimate the effective absorption cross section $\sigma$ from our results, letting $\dot{M}=\sigma \rho v_0$. For simulation \texttt{IVB} one finds $\sigma\sim 400$ at the peak of accretion, corresponding to an absorbing disk with radius 2 orders of magnitude larger than the BH scale \cite{Unruh:1976fm}. We do not understand the origin of such large accretion rates, but we suspect they are related to a transient stage. Unfortunately, our setup does not allow us to simulate more extreme length scales; hence, we never see past the transient and into the stationary regime.

For simulations with smaller BH masses, the residual scalar field outside the BH increases by several orders of magnitude as seen in Table~\ref{table:simulations_results}, although it is still but a small fraction of the total energy. One can expect that for even smaller BH masses the BH can pierce through without "eating" the entire BS. Unfortunately, as we said, in order to evolve such a small BH, one would need too large an amount of computational resources.

\subsection{Black hole motion and tidal capture\label{subsec:tidal}}
%
\begin{table*}[htb]
\caption{Summary of the results of the dynamical evolution of the initial data in Table~\ref{table:simulations1}. Here, $M_f$ is the final BH irreducible mass, and $v_f$ is the final BH velocity as calculated from the puncture trajectory; in parentheses we show the expected value $M_{\rm BH} v_0/M_{\rm tot}$ from momentum conservation, assuming that the entire BS is accreted onto the BH (notice the good agreement between these two estimates). Note that $E^{\rm rad}$ and $P^{\rm rad}$ are the energy and momentum radiated in GWs, respectively. They are calculated from $\psi_4$ and compared to a Newtonian, quadrupolar approximation which includes no accretion (number in parentheses, see main text for further details).
Finally the total momentum and energy flux of the {\it scalar} field into the BH horizon are the last two entries. The junk radiation exists in all cases and we neglect it.
\label{table:simulations_results}}
\begin{ruledtabular}
\begin{tabular}{l@{\hskip 0.1in}c@{\hskip 0.1in}c@{\hskip 0.1in}c@{\hskip 0.1in}c@{\hskip 0.1in}c@{\hskip 0.1in}c@{\hskip 0.1in}c@{\hskip 0.1in}c@{\hskip 0.1in}c@{\hskip 0.1in}c@{\hskip 0.1in}}
Run          & $M_{\rm BH}$&$M_f$ & $v_0$         & $v_f$                       & $10^4 E^{\rm rad}$ & $10^4 P_z^{\rm rad}$ &$Q_{t}^{\rm initial}$ & $Q_{t}^{\rm final}$ & $Q_{z}^{\rm initial}$ & $Q_{z}^{\rm final}$\\
\hline
\texttt{IA}  &  $1.0$        &     $1.56$ & $10^{-4}$     & $-2.9 \times 10^{-4}$ (0) & $2.9$ (11.7)       &  $-0.24$ & $0.53$ & $7.7 \times 10^{-5}$ & $0$ & $4.1 \times 10^{-8}$\\
\texttt{IB}  &  $1.0$        &    $1.64$  & $0.5$         & $0.33$ (0.30)               & $12.5$ (35.6)     & $4.1$ & $0.54$ & $3.9\times 10^{-4}$ & $0$ & $-7.4 \times 10^{-5}$\\
\texttt{IIA} & $0.4$       &    $0.94$  & $10^{-4}$     & $-1.6 \times 10^{-3}$ (0) & $0.63$ (1.9)      &  $6.5\times 10^{-3}$ & $0.52$ & $2.6 \times 10^{-3}$ & $0$ & $-3.1\times 10^{-6}$\\
\texttt{IIB} & $0.4$       &    $1.01$  & $0.5$         & $0.20$ (0.20)                &  $5.5$ (5.7)      &  $1.5$ & $0.54$ & $4.0 \times 10^{-3}$ & $0$ & $-6.4 \times  10^{-4}$ \\
\texttt{IIIA} & $0.2$      &    $0.73$  & $10^{-4}$     & $-2.5 \times 10^{-3}$ (0) & $0.20$ (0.47)     & $9.0\times 10^{-3}$ & $0.51$ & $4.2 \times 10^{-3}$ & $0$ & $-1.6 \times 10^{-4}$ \\
\texttt{IIIB} & $0.2$      &    $0.78$  & $0.5$         & $0.12$ (0.13)               &  $2.2$ (1.4)     &  $0.47$ & $0.54$ & $1.1 \times 10^{-2}$ & $0$ & $-1.3 \times 10^{-3}$\\
\texttt{IVA} & $0.1$       &    $0.61$  & $10^{-4}$     & $-7.1 \times 10^{-3}$ (0) & $0.10$ (0.12)   & $4.7 \times 10^{-3}$ & $0.51$ & $1.5 \times 10^{-2}$ & $0$ & $-1.0 \times 10^{-3}$ \\
\texttt{IVB} & $0.1$        &    $0.64$     & $0.5$             &  $0.064$ (0.077)    &  $0.75$ (0.36)     &  $0.11$ & $0.53$ & $2.2 \times 10^{-2}$ & $0$ & $-1.2 \times 10^{-3}$\\
\end{tabular}
\end{ruledtabular}
\end{table*}
To interpret our results and to compare them with simple estimates,
consider first the motion, in Newtonian dynamics, of two pointlike particles of mass $m_1$ at $z=z_1(t)$ and $m_2$ at $z=z_2(t)$. We thus neglect tidal effects (although they are apparent in our results, cf. Fig.~\ref{fig:snapshotsIB}) and relativistic effects. Defining $d=z_2-z_1$ and using Newtonian dynamics (or equivalently, energy and momentum conservation), one finds
\be
\ddot{d}=-\frac{m_1+m_2}{d^2}\,,\qquad \ddot{z}_1=\frac{m_2}{d^2}\,.
\ee
For our system, we take $m_1=M_{\rm BH},\, m_2=M$, and we integrate them with initial conditions appropriate for the initial data. For \texttt{IB}, $z_1(0)=-200,\, d=200$. We find that the BH and BS are separated by a distance $d=5$ when the BS sits at $z_2=-7.7$, which is consistent with the left bottom panel in Fig.~\ref{fig:snapshotsIB}.

\begin{figure}[htpb]
  \centering
  \includegraphics[width=0.52\textwidth]{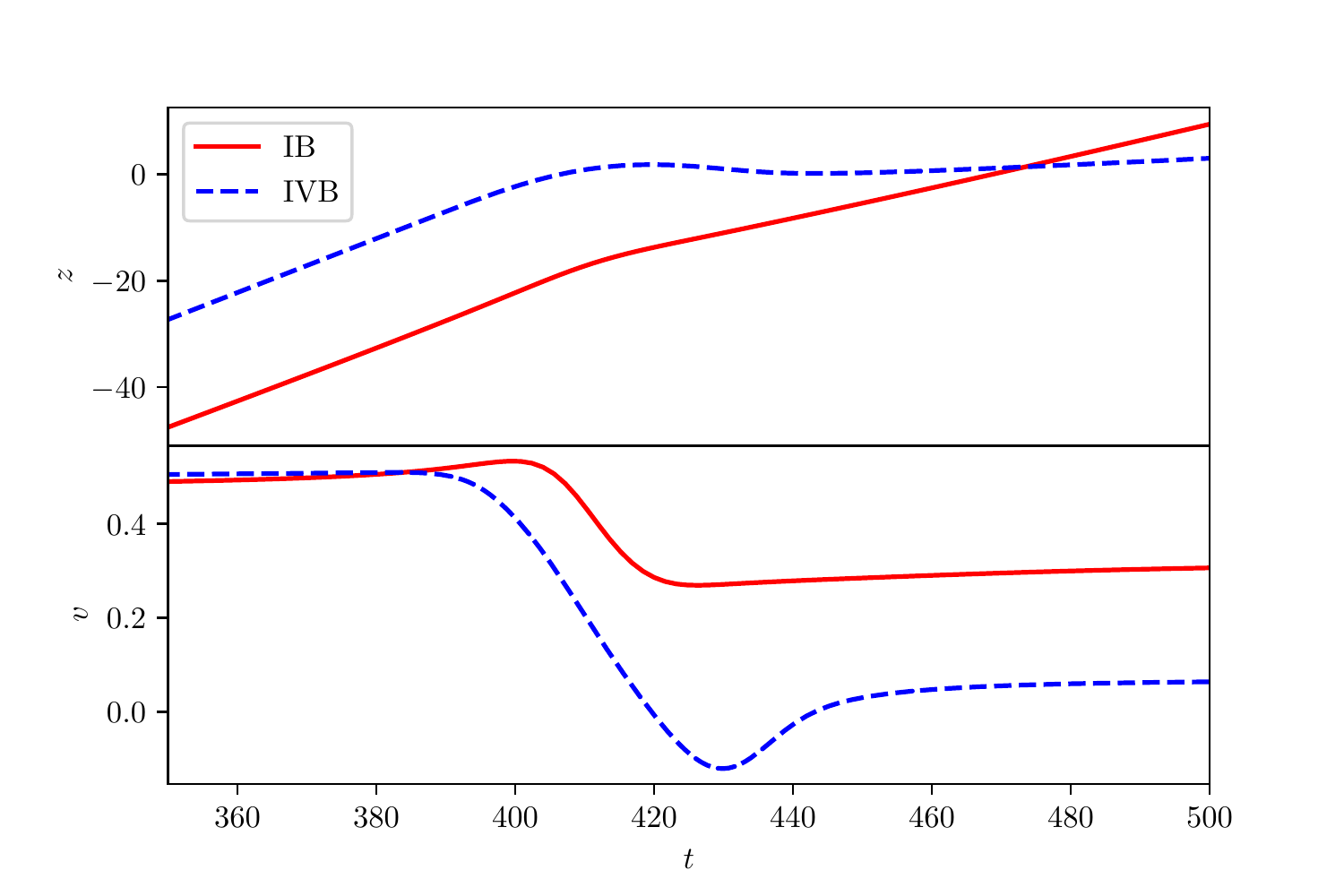}
  \caption{Location $z$ and the speed $v$ of the BH for \texttt{IB} and \texttt{IVB}, as read from the puncture location. The interaction between the BH and the BS is clear from these 
	data, and it translates into a deceleration starting at $t\sim 400$ and lasting for $20$, roughly the time taken to cross the bulk of the BS. Notice also that the BH velocity is {\it negative}
	for a small amount of time in simulation \texttt{IVB}: the BH is tidally captured by the boson star.
		}
  \label{fig:bh_loc_v}
\end{figure}
The BH velocity can be estimated from our numerical results, using the puncture trajectory. These estimates are shown in Fig.~\ref{fig:bh_loc_v} for simulations \texttt{IB} and \texttt{IVB},
and also in Table~\ref{table:simulations_results}. Due to our initial gauge condition $\beta = 0$, the puncture velocity is zero at the initial time, but it will reach $v_0$ over a period of time (see Appendix~\ref{app:boosted_bh_gauge}). Notice that the BH velocity after interaction with the BS agrees well with simple momentum-conservation estimates.
For extreme length ratios, the agreement is not as good due to the extreme requirements necessary; our simulation is not yet accurate enough, and we would require larger resolutions to fully capture the dynamics to the necessary level of precision.

Nevertheless, there are interesting details in how the asymptotic velocities are achieved, which require a deeper analysis. We find that the BH pierces through the boson star and produces a tidal  ``stretching,'' with a significant amount of energy deposited in such a configuration. A point of near-maximum distortion is shown in the lower left panel of Fig.~\ref{fig:snapshotsIVB}. At this instant, the velocity of the BH is close to zero, and the tidally distorted cloud is slowly moving in the positive $z$ direction. Thus, the cloud pulls the BH, which then acquires a velocity in the negative $z$ direction, as seen in Fig.~\ref{fig:bh_loc_v}. The pull is significant and can make the BH velocity large, but negative. As the cloud relaxes from this tidal pull, it is accreted by the BH and transfers all its positive momentum to the BH, which then reaches its asymptotic value, consistent with momentum conservation.
This momentarily retrograde motion is directly responsible for the different accretion peaks in Fig.~\ref{fig:bh_mass}, and for the features in GW emission that we discuss below.

In other words, what we find is an extreme example of ``tidal capture''~\cite{10.1093mnras172.1.15P,1977ApJ...213..183P}, which in our setup leads to the subsequent accretion of the entire boson star. As discussed in Ref.~\cite{10.1093mnras172.1.15P}, the maximum tidal oblateness that can be induced is of order $M_{\rm BH}/M$ for a BH sitting at the BS ``radius.'' The associated tidal energy is $\sim M_{\rm BH}^2/R_{98}$. Since the BH mass grows by the time it reaches the other BS extreme, this quantity can be a sizable fraction of the initial kinetic energy, leading to the stoppage of the BH. To prevent the BH backward motion, smaller BH masses need to be considered, or more dilute boson stars. Unfortunately, either choice requires more extreme length ratios, which we are unable to study at this point.

One of our original motivations was to study dynamical friction in a full nonlinear realistic setup.
Dynamical friction is of order $4\pi M\rho/v^2$ for a constant density boson star~\cite{Hui:2016ltb,Lancaster:2019mde,Annulli:2020lyc,Vicente:2022ivh}. We can estimate when dynamical friction is dominant with respect to gravitational acceleration; both effects will impart similar accelerations (albeit of different sign) when
\be
\frac{4}{3}\pi\rho r\sim \frac{4\pi M_{\rm BH}\rho}{v^2}\,,
\ee
where energy conservation requires that the BH velocity, in the constant-BH-mass approximation, satisfies $v^2=v_R^2+4\pi\rho/3(R_{98}^2-r^2)$. Here $v_R$ is the BH velocity when it reaches the BS radius.
Even in the most favorable situation when the BH starts from rest at infinity, $v_R^2=2M/R_{98}$, and therefore we find that dynamical friction dominates only at distances
\be
r\lesssim \gamma R_{98}\,,
\ee
where $\gamma$ is defined by $M_{\rm BH}=\gamma M$. Thus, we see how challenging it is to perform simulations of self-gravitating objects where dynamical friction can be isolated.

\subsection{Gravitational-wave emission}

When the BS is Newtonian and the BH velocity is nonrelativistic, we can use the quadrupole approximation to estimate waveforms and radiated fluxes.
In addition, when the BH mass $M_{\rm BH}$ is much smaller than that of the BS mass $M$, one can find simple solutions to the dynamics of the system and consequent GW emission.
In such a setup, the BH simply follows a spacetime geodesic parametrized by a radial position $r(t)$, in a background whose geometry is dictated by a BS fixed at the center of coordinates. The quadrupole approximation yields
\be
\frac{dE}{dt}=\frac{8}{15}M_{\rm BH}^2\left(3\dot{r}\ddot{r}+r\dddot{r}\right)^2\,.
\label{eq:quad}
\ee
We first approximate the BS as a scalar blob of constant density and radius $R_{98}$, with vacuum outside. In this Newtonian setup the motion of the BH in the exterior can be calculated using energy conservation,
\be
\dot{r}^2-2M/r=v_0^2\,,\label{app:motion_exterior}
\ee
where $v_0$ is the asymptotic velocity at large distances. In the interior, one needs an accurate description of the BS gravitational potential.
In the approximation of constant density $\rho=3M/(4\pi R^3)$, one finds
\be
\dot{r}^{2}+\frac{4}{3}\pi\rho\left(r^{2}-R_{98}^{2}\right)=v_{R}^{2}
\ee
in the interior, where $R_{98}$ is the BS radius and $v_R$ the velocity at $R_{98}$ which can be obtained from \eqref{app:motion_exterior}.

Consider first infalls from rest. Integrating the quadrupole formula in the exterior, one then finds (using $dt=dr/\dot{r}$ to perform a radial integration)
\be
E_{\rm rad}^{\rm ext}=\frac{16\sqrt{2}}{105}\left(\frac{M}{R_{98}}\right)^{5/2}\frac{M_{\rm BH}^2}{R_{98}}\,.
\ee
In the interior one finds instead
\be
E_{\rm rad}^{\rm int}=\left(\frac{63 \cot ^{-1}\left(\sqrt{2}\right)}{\sqrt{2}} -7\right) E_{\rm rad}^{\rm ext}\,.
\ee

For finite velocity or realistic matter density, a numerical integration of the equation of motion, along with that of the quadrupole formula is necessary.
The result of such an integration is shown in Table~\ref{table:simulations_results}, where we use the fully relativistic boson star solution, and a slow motion approximation.
We truncate the integration at the origin when $v_0=0$,
but we assume that the BH plunges through and emerges on the other side for $v_0=0.5$.

\begin{figure}[htpb]
  \includegraphics[width=0.52\textwidth]{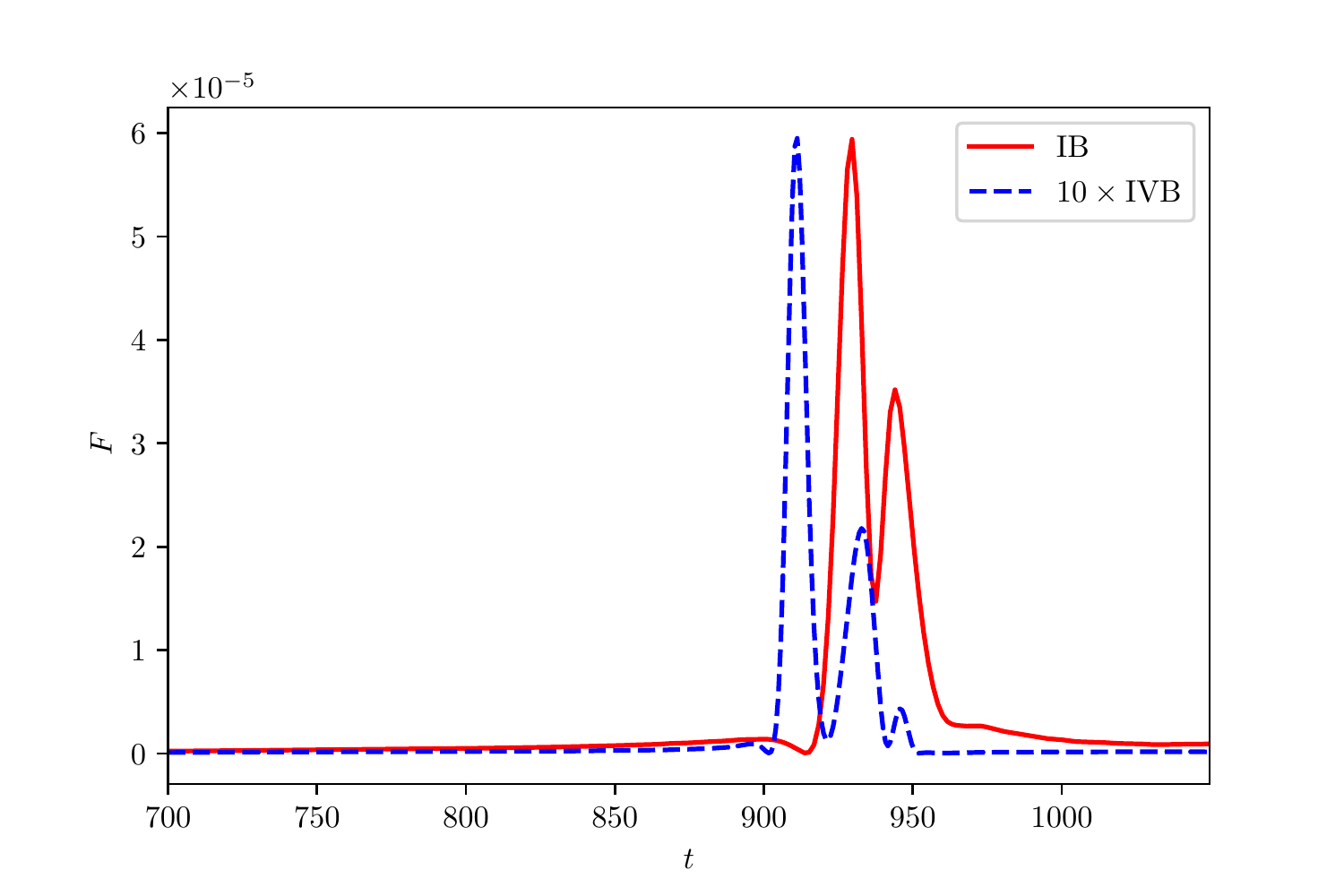}
  \caption{Energy flux $F = \frac{d E^{\mathrm{rad}}(t)}{dt}$ of the GW, which is the integration of $\Psi_4$ on the sphere $r = 500$. The blue dashed line is multiplied by $10$.}
  \label{fig:gw_flux_500}
\end{figure}
Our fully relativistic results for GW emission are shown in Table~\ref{table:simulations_results} and Fig.~\ref{fig:gw_flux_500}.
We compute the energy and momentum flux of GWs at $r = 500$. There is a pulse of ``junk'' radiation in the initial data, which is easily distinguished from the physical pulse~\cite{Sperhake:2008ga}. Hence we are able to discard nonphysical contributions to the total radiated fluxes and energies. As seen in Table~\ref{table:simulations_results}, our Newtonian estimate is in good agreement with the full relativistic results, especially if one considers how simple the Newtonian model is, with no accretion included. The flux of energy in GWs for extreme length ratios, Fig.~\ref{fig:gw_flux_500},
shows three emission peaks, which are caused by the BH tidally induced retrograde motion and accretion and momentum-induced forward motion described above.

Table~\ref{table:simulations_results} also shows the linear momentum carried by GWs. Since radiation is forward focused, linear momentum in GWs is along the direction of the BH motion, in the positive $z$ direction. The only exception concerns simulation \texttt{IA}, for which the collision happens from rest and the BS is less massive than the BH. It is thus expected that this is best described by a BS falling onto a BH, hence a negative total linear momentum radiated. In any case, the linear momentum carried by GWs is too small to influence any of the dynamics of the system.

\subsection{Late-time decay of the scalar}
%
\begin{figure*}[thpb]
  \centering
  \subfloat[][]{\includegraphics[width=0.52\textwidth]{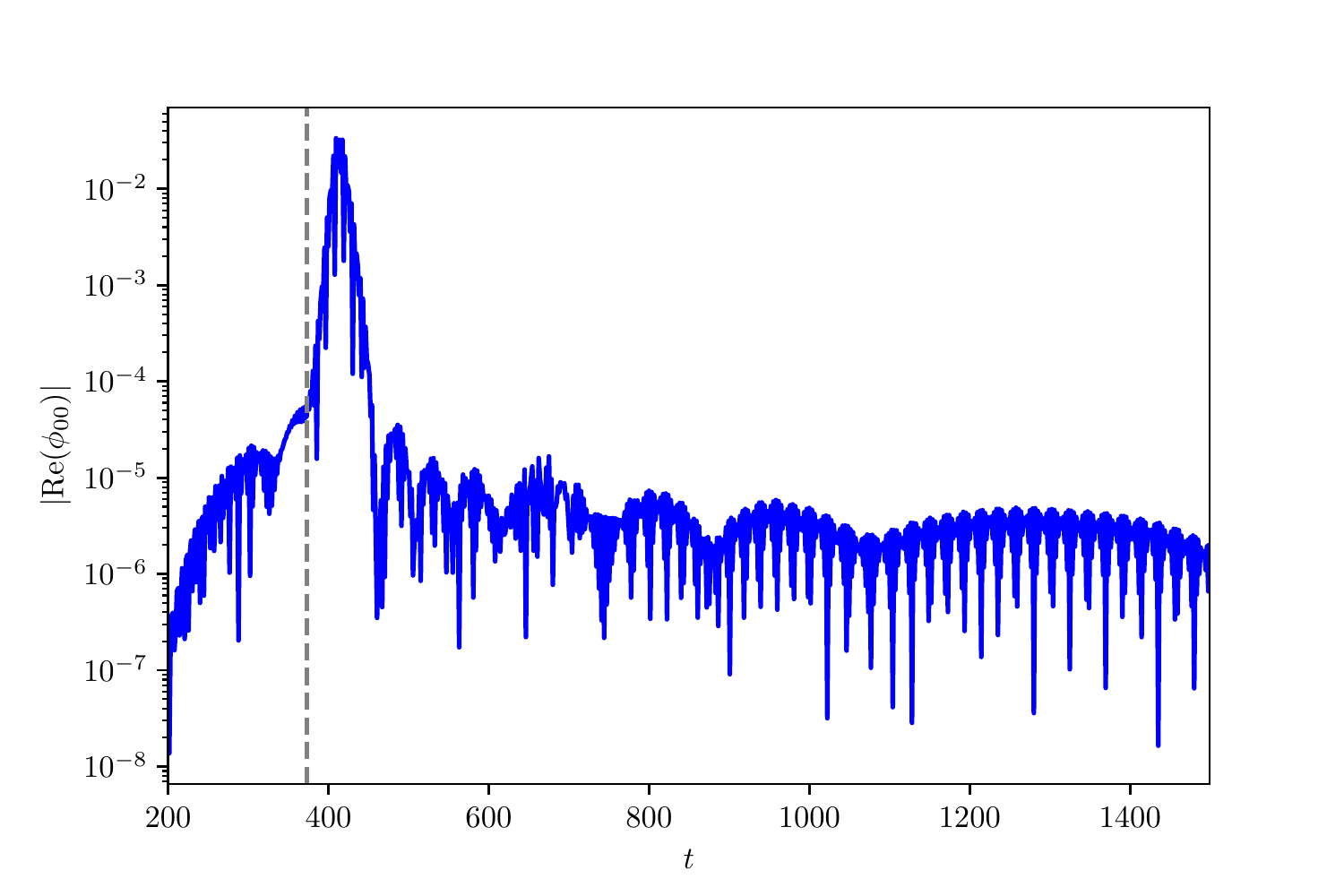}}
  \subfloat[][]{\includegraphics[width=0.52\textwidth]{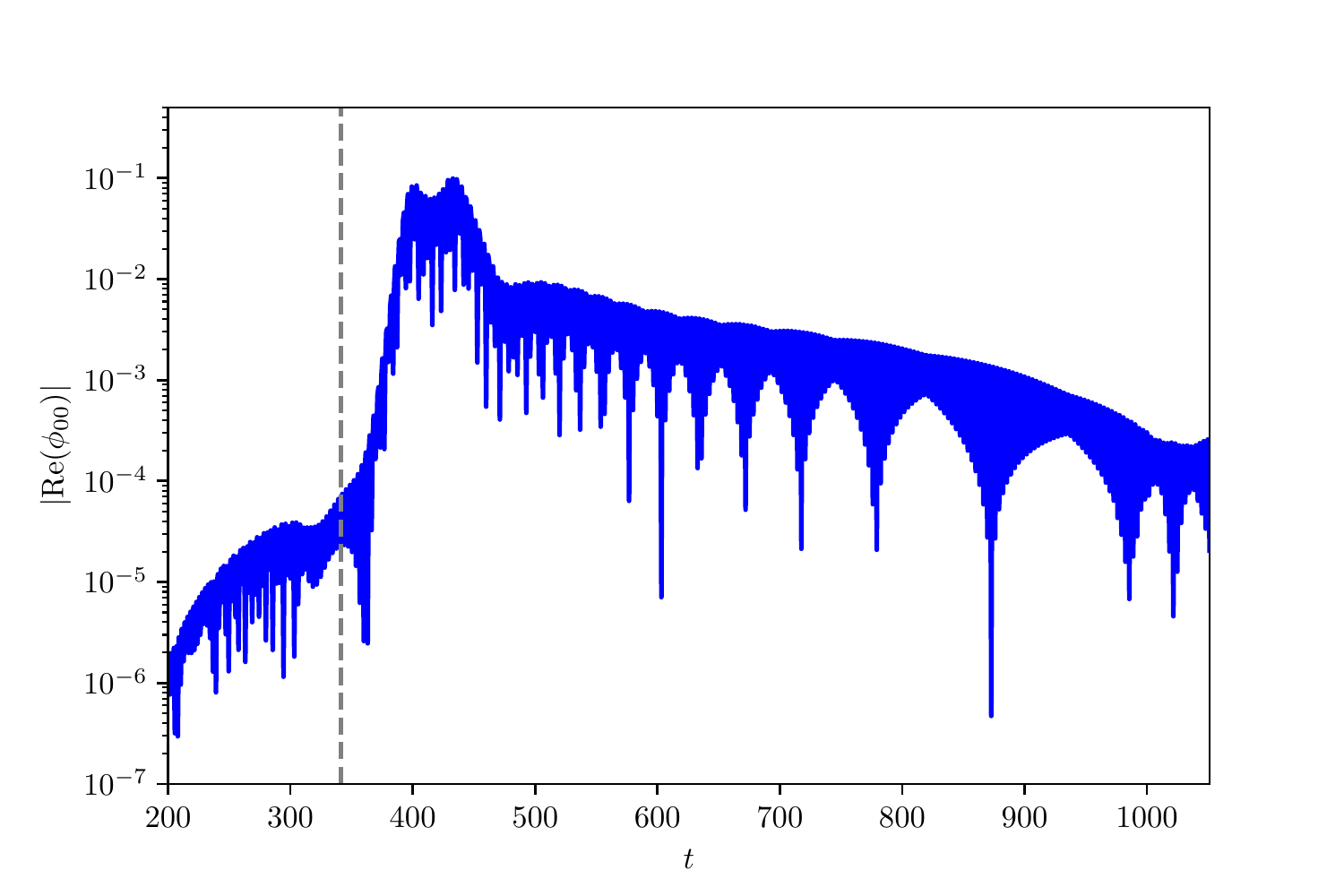}}
  \caption{Real part of the $l=m=0$ multipole of the scalar field on sphere $r = r_{\mathrm{BH}} + 1$ around the BH for \texttt{IB} and \texttt{IVB}, where the origin is the position of the BH and $r_{\mathrm{BH}}$ is the BH horizon radius. The dashed gray line indicates a ``merger instant,'' (somewhat arbitrarily) defined as the instant where $\abs{\phi_{00}} > 10^{-2}$ on the sphere $r = r_{\mathrm{BH}} + 1$. The monopolar component grows once the BH plunges into the BS, after which we see an exponentially decaying stage of a rate ($\omega_I\sim -0.15$ for \texttt{IB} and $\omega_I\sim -0.075$ for \texttt{IVB}, in rough agreement with expectations from the quasi-bound state calculation within a linearized approach). Our calculations indicate that this is the first overtone and that a ``gravitational atom'' was created. At late times, we see a power-law decay, $\phi_{00}\sim t^{-1.5}$ for simulations \texttt{IB}, as expected for massive fields~\cite{Koyama:2001ee,Witek:2013sup}. For simulation \texttt{IVB} we do not have clear control on the very-late-time behavior: this would require a longer simulation, which for these extreme values is very challenging to achieve, while keeping precision under control.
	\label{fig:bound_state}}
\end{figure*}
The BH accretes most of the BS material in a violent accretion stage, as we saw. However, a small fraction of the BS leftovers linger around the BH in a quasi-bound state co-moving with the BH, as might be expected for massive scalars. To understand if such states correspond to those in perturbation theory~\cite{Brito:2015oca}, we use a spherical harmonic decomposition in the BH frame, as explained in Sec.~\ref{subsec:diagnostics}.

Figure~\ref{fig:bound_state} shows the monopolar ($l=m=0$) component of the scalar field on a sphere around the BH extracted close to the BH for simulations \texttt{IB} and \texttt{IVB}. The field oscillates at a frequency $\mu$, as expected for massive fields. We also see a clear exponential decay of the field after the collision, indicating that the system is relaxing in one of its characteristic modes of vibration, corresponding to the so-called ``gravitational atom'' (with a BH ``nucleus'' and the light scalar field as the ``electron'')~\cite{Detweiler:1980uk,Brito:2015oca,Baumann:2019eav}.

To test this picture and to compare with the decay timescale of a bound state within perturbation theory, we calculate the quasi-bound states of massive scalars in a fixed Schwarzschild geometry, using Leaver's continued fraction method~\cite{Leaver:1985ax,Brito:2015oca}. In other words, the field is assumed to be quasistationary $\phi\sim e^{i\omega t}$ and evolving on a fixed BH background.
The characteristic frequencies $\omega=\omega_R+i\omega_I$. For simulation \texttt{IA}, for example the final BH mass $M_f \sim 1.56$, and the decay rate seen in the time evolution is close to that predicted by mode calculations, at $\omega_I=-0.13$. For simulations \texttt{IB} and \texttt{IVB}, we find a similar behavior, although the agreement with linearized mode calculations is not as good, nor as clean, since the BH is still accreting at late times (which changes its mass, and the quasi-bound state frequencies are very sensitive to the precise value of the BH mass).

The quasi-bound states decay exponentially, and eventually may give way to power-law tails. For massive scalars, the tail stage is dominated at intermediate times by a behavior $\phi_{00}\sim t^{-3/2}$
for spherically symmetric modes~\cite{Koyama:2001ee,Witek:2013sup}. This behavior is independent of the presence of the BH and is already present in Minkowski: it is a characteristic of massive fields.
Our results for the late-time behavior, seen for example in Fig.~\ref{fig:bound_state}, are consistent with a decay close to the theoretical power-law tail $t^{-3/2}$, characteristic of intermediate times~\cite{Koyama:2001ee,Witek:2013sup}. We note that this behavior is consistent with what we see in runs \texttt{IA}, \texttt{IVA},  and \texttt{IB}. For simulation \texttt{IVB} we do not have any indication that the field settled to its late-time value yet (by the time the simulation finishes, the BH is still accreting).
At very late times, one expects a slightly different behavior, $\phi_{00}\sim t^{-5/6}$, for which we have no evidence. There are several possible reasons for this, one of them being that we are not able to evolve the spacetime long enough for this decay to dominate.

\section{Conclusion}
We have performed some of the most challenging simulations to date involving BHs and BSs, with length ratios as large as $\sim 62$.
Our goal was to understand how bosonic structures---which could describe dark matter---interact with BHs, and which dynamical friction or accretion they induce on the black holes.
Our fully relativistic results are in good agreement with Newtonian estimates for the motion and asymptotic velocities of the objects,
as well as for GW emission. To our surprise, even at the most extreme length ratios we considered, the boson star is entirely accreted by the BH. The reason, we believe, is that the BH is tidally captured by the boson star. This seems to be an extreme case of tidal capture not reported previously.
At late times, a ``gravitational atom'' is formed, where a massive  BH is surrounded by a quasi-bound state of the scalar field, the BS remnant.

One of our goals was to see the fate of a boson star which had been pierced by a high-velocity BH, and how dynamical friction on the BH compares to estimates in the literature.
Unfortunately, because of tidal capture, the BS is swollen and dynamical friction turns out to be strongly enhanced due to transient accretion. We estimate in Sec.~\ref{subsec:tidal}
that very challenging simulations need to be done, if tidal friction is to be identified at the full nonlinear level in self-gravitating structures.

In our configuration, the BS is kept fixed in all scenarios, but we believe the results would be similar for other BS parameters.
We focus on spherically symmetric, neutral configurations, but there are obviously new phenomena when one extends the initial data to spinning or charged configurations (see also
Ref.~\cite{Sanchis-Gual:2020mzb} for a possible similar end state with spin). We are currently starting these studies.

\begin{acknowledgments}
We are grateful to Ulrich Sperhake for sharing with us a script to evaluate gravitational-wave related physical quantities.
The data analysis in this project is carried out using the kuibit Python package \cite{kuibit}.
Z.Z.\ acknowledges financial support from China Scholarship Council (No.~202106040037).
V.C.\ is a Villum Investigator and a DNRF Chair, supported by VILLUM FONDEN (grant no.~37766) and by the Danish Research Foundation. V.C.\ acknowledges financial support provided under the European
Union's H2020 ERC Advanced Grant ``Black holes: gravitational engines of discovery'' grant agreement
no.\ Gravitas–101052587.
T.I.\ acknowledges financial support provided under the European Union's H2020 ERC, Starting
Grant agreement no.~DarkGRA--757480.
M.Z.\ acknowledges financial support provided by FCT/Portugal through the IF programme, grant IF/00729/2015, and
by the Center for Research and Development in Mathematics and Applications (CIDMA) through the Portuguese Foundation for Science and Technology (FCT -- Funda\c{c}\~ao para a Ci\^encia e a Tecnologia), references UIDB/04106/2020, UIDP/04106/2020 and the projects PTDC/FIS-AST/3041/2020 and CERN/FIS-PAR/0024/2021.
This project has received funding from the European Union's Horizon 2020 research and innovation programme under the Marie Sklodowska-Curie grant agreement No 101007855.
We thank FCT for financial support through Project~No.~UIDB/00099/2020.
We acknowledge financial support provided by FCT/Portugal through grants PTDC/MAT-APL/30043/2017 and PTDC/FIS-AST/7002/2020.
The results of this research have been achieved using the DECI resource Snellius based in The Netherlands at SURF with support from the PRACE aisbl, and
the Navigator cluster, operated by LCA-UCoimbra, through project~2021.09676.CPCA.

\end{acknowledgments}

\appendix

\section{Numerical convergence \label{app:convergence}}

To check the convergence of our numerical results, we define the usual convergence factor
\begin{equation}
Q_{n}=\frac{f_{\Delta_{c}}-f_{\Delta_{m}}}{f_{\Delta_{m}}-f_{\Delta_{h}}}=\frac{\Delta_{c}^{n}-\Delta_{m}^{n}}{\Delta_{m}^{n}-\Delta_{h}^{n}}
\end{equation}
where $n$ is the order of the finite difference scheme used, and
$f_{\Delta_{c}}$, $f_{\Delta_{m}}$, and $f_{\Delta_{h}}$ are the numerical solutions obtained for a given function $f$ at resolutions $\Delta_c$, $\Delta_h$ and $\Delta_h$, respectively.
For these tests
we performed simulations for configuration \texttt{IB} (see Table~\ref{table:simulations1}) with resolutions  $\Delta_c = 2.4 M$, $\Delta_m = 2 M$ and $\Delta_h = 1.6 M$.

\begin{figure}[!htbp]
  \centering
  \includegraphics[width=0.45\textwidth]{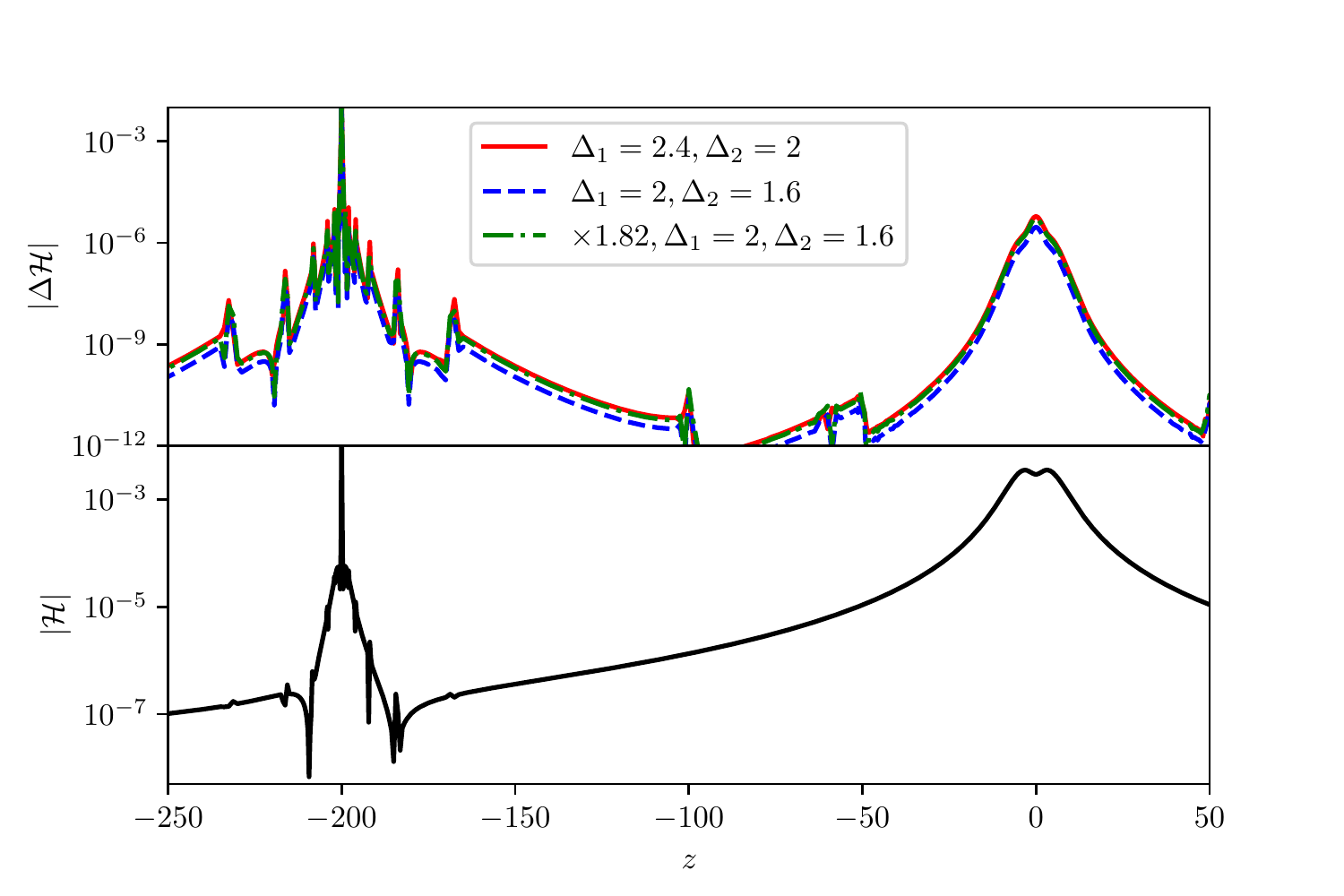}
  \caption{\textbf{Top:} convergence of the Hamiltonian constraint violation at $t = 0$ for \texttt{IB}. The green line is multiplied by $Q_4 = 1.82$, the expected factor for fourth-order convergence. \textbf{Bottom:} Richardson extrapolation used to obtain the value of the Hamiltonian constraint as $\Delta \to 0$.\label{hc_cov_1b}}
\end{figure}
As explained in the main text, we use a simple superposition procedure to build initial data.
The Hamiltonian constraint is therefore expected to converge to a small, but nonzero, value. This is shown in Fig.~\ref{hc_cov_1b}, where a fourth-order convergence can be seen.
We experimented with other methods for the superposition operation, but saw no noticeable advantage. 
For the future, we intend to try the method of Ref.~\cite{Helfer:2021brt}, or a similar method, to check if any improvement is observed.

\begin{figure}[!htbp]
  \centering
  \includegraphics[width=0.45\textwidth]{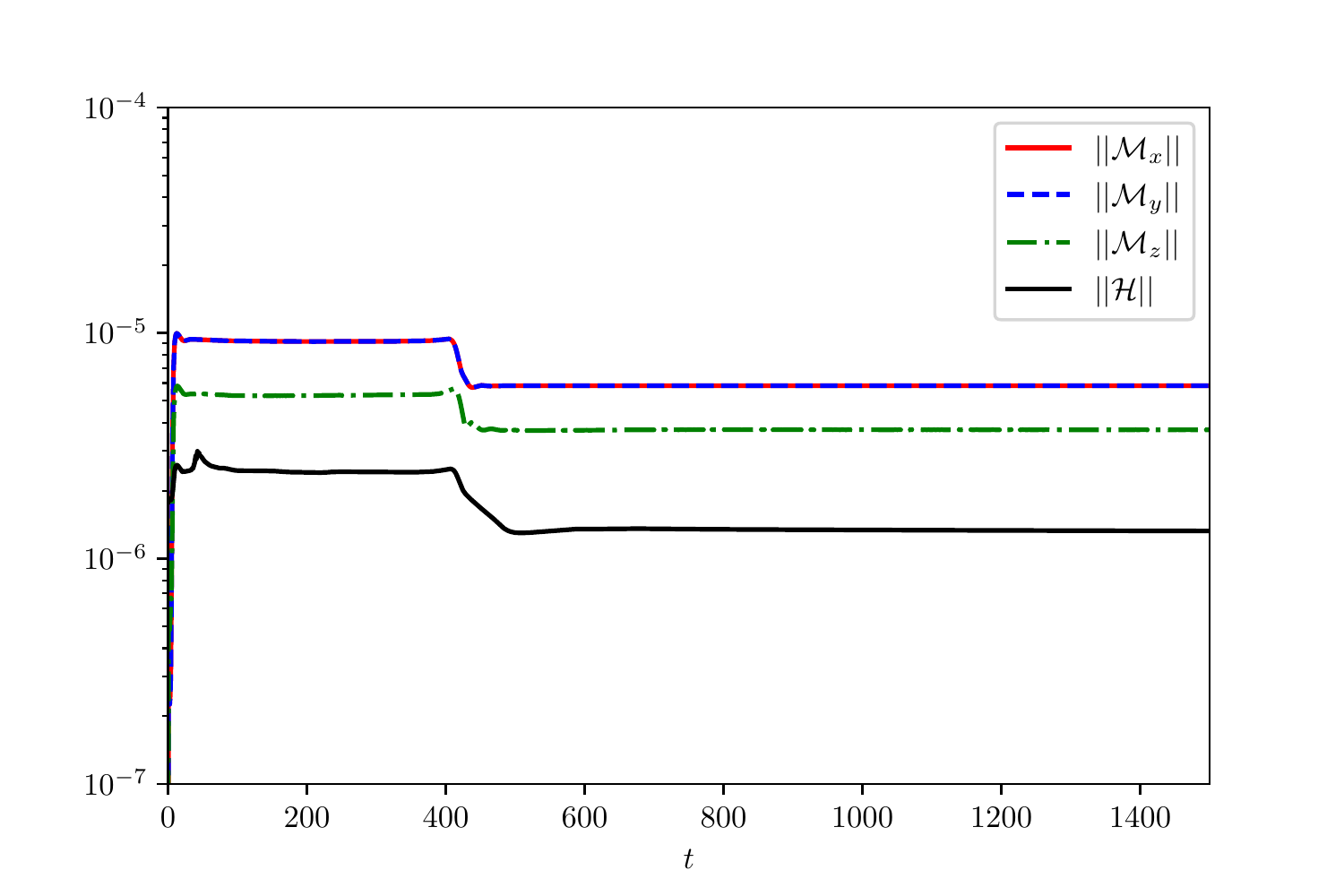}
  \caption{Violation of the Hamiltonian and momentum constraints as functions of time for run \texttt{IB}.
    \label{fig:bh_bs_4b_constraits_norm2}}
\end{figure}
We also monitor the $\ell^2$-norm of the constraint violations during the evolution (see Fig.~\ref{fig:bh_bs_4b_constraits_norm2}) to confirm these do not grow with time.

\begin{figure}[!htbp]
 \centering
 \includegraphics[width=0.45\textwidth]{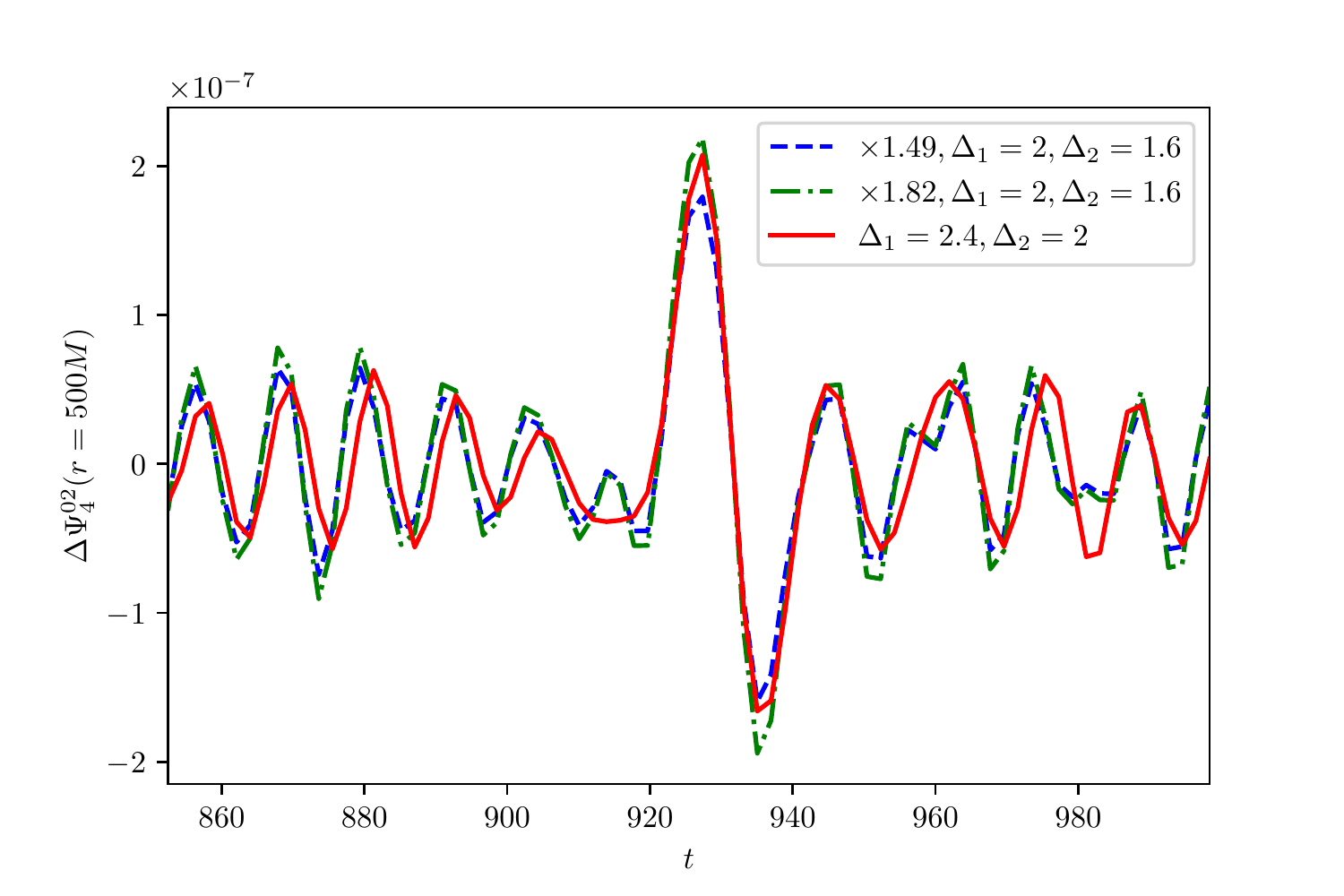}
  \caption{Convergence analysis of the $l=0$, $m=2$ multipole of $\Psi_{4}$, extracted at $r = 500M$. The blue line shows the expected result for third-order convergence ($Q_3 = 1.49$), while the green line shows the expected result for fourth-order convergence ($Q_4 = 1.82$).
  \label{fig:cov}}
\end{figure}
Finally, we plot in Fig.~\ref{fig:cov} the convergence analysis for the $l=0$,
$m=2$ multipole of $\Psi_{4}$, extracted at $r = 500M$, for configuration
\texttt{IB}. The results are compatible with a convergence order between third and fourth order.

\section{Boosted BH simulation}\label{app:boosted_bh_gauge}

\begin{figure}[!htbp]
  \includegraphics[width=0.5\textwidth]{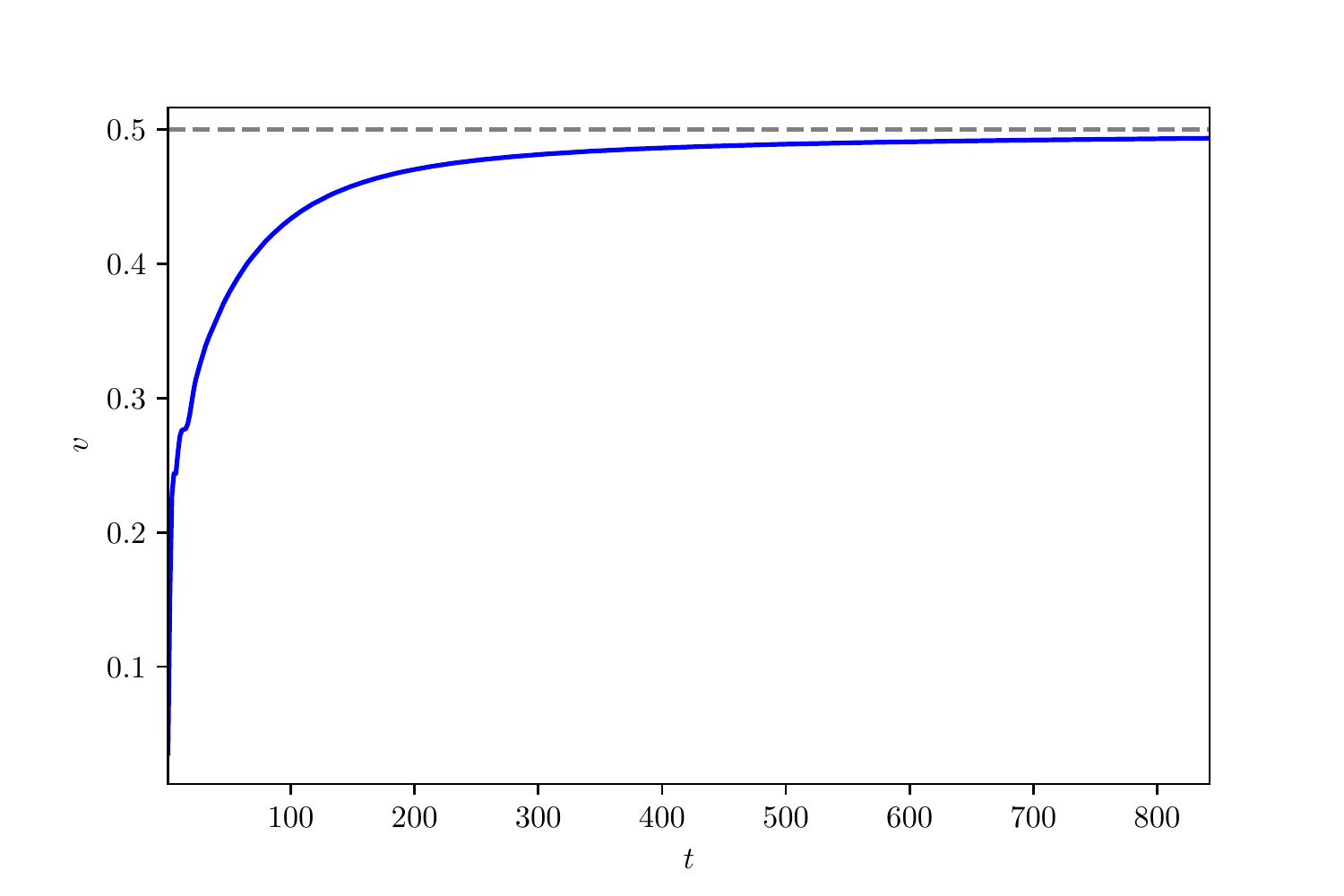}
  \caption{Puncture velocity of a single boosted BH with $M = 1$ and $v = 0.5$. The dashed gray line is $v = 0.5$.}
  \label{fig:boosted_bh_gauge}
\end{figure}

Since we set $\beta = 0$ at the initial time, the puncture speed is (initially) zero. Here we discuss the effect of this initial gauge condition.
Figure~\ref{fig:boosted_bh_gauge} shows a simulation of an isolated boosted BH with $M = 1$ and $v = 0.5$. One can see that the puncture speed (as measured by the zero of the shift vector $\beta$) of the boosted BH approaches the speed of BH $v = 0.5$ after a period of time and eventually stays at $v = 0.5$. This means that the instantaneous puncture speed before the collision can be used to gauge the speed of the BH.

\bibliography{references} 

\end{document}